\documentclass[a4paper,11pt,bookmarks=true]{article}
\pdfoutput=1
\usepackage{jheppub}
\usepackage[colorlinks=true,linkcolor=blue,citecolor=blue, urlcolor=blue]{hyperref}  
\usepackage[dvipsnames]{xcolor}
\usepackage{graphicx}
\usepackage{float}
\usepackage{placeins}
\usepackage{dcolumn}
\usepackage{bm}
\usepackage{amsmath}
\usepackage{enumitem}
\usepackage[capitalise]{cleveref}
\usepackage[normalem]{ulem} 
\usepackage{subcaption}
\usepackage{multirow}

\def\dd{\mathrm{d}}

\begin{document}

\title{No-quenching baseline for energy loss signals in oxygen-oxygen collisions}

\author{Jannis Gebhard,}%
\author{Aleksas Mazeliauskas}
\author{and Adam Takacs}
\affiliation{Institute for Theoretical Physics, University of Heidelberg, Philosophenweg 12, 69120 Heidelberg, Germany}

\emailAdd{gebhard@thphys.uni-heidelberg.de}
\emailAdd{a.mazeliauskas@thphys.uni-heidelberg.de}
\emailAdd{takacs@thphys.uni-heidelberg.de}

\date{\today}

\abstract{In this work, we perform computations of inclusive jet, and semi-inclusive jet-hadron cross sections for minimum bias oxygen-oxygen collisions at RHIC and LHC collision energies. We compute the no-quenching baseline for the jet nuclear modification factor $R_\mathrm{AA}$ and jet-, and hadron-triggered semi-inclusive nuclear modification factors $I_\mathrm{AA}$. We do this with state-of-the-art nuclear parton distribution functions (nPDFs), next-to-leading-order matrix elements, parton shower, and hadronization. We observe deviations from unity due to cold-nuclear matter effects, even without quenching. We demonstrate that the parton distribution uncertainties constitute a significant obstacle in detecting energy loss in small collision systems. Hadron-triggered observables are particularly sensitive to uncertainties due to correlations between the trigger and analyzed particles. For jet-triggered $I_\mathrm{AA}$, there exists a kinematic window in which nPDF and scale uncertainties cancel dramatically while showing little sensitivity to parton shower and hadronization models, addressing a major limiting factor for energy loss discovery in small systems.}

\maketitle

\section{Introduction}
The ultra-relativistic collisions of lead nuclei at the Large Hadron Collider (LHC) at CERN and gold nuclei at the Relativistic Heavy Ion Collider (RHIC) at BNL compress matter to energy densities that far exceeds the rest density of nuclear matter. In such extreme conditions, the partonic degrees of freedom become deconfined and a new form of matter known as Quark-Gluon Plasma (QGP) is created. The presence of the QGP is inferred from a large number of hadronic and electromagnetic observables that are significantly modified from more elementary proton-proton collisions. 
The prime example is the collective flow of produced particles in the transverse plane of the collision~\cite{Ollitrault:1992bk,Heinz:2013th,Gale:2013da,Busza:2018rrf}. Thanks to the unequal transverse density gradients in the collision plane, the expanding QGP creates anisotropic particle flow. 
Another phenomenon is the suppression (quenching) of high momentum hadrons and jets as they interact and deflect during their propagation in the QGP~\cite{Bjorken:1982tu,Gyulassy:1990ye,Wang:1992qdg,Qin:2015srf,Connors:2017ptx}. 
Surprisingly, collective flow signals have been discovered even in small collision systems, namely peripheral nuclear collisions, proton-nucleus, and proton-proton collisions~\cite{CMS:2010ifv,Schenke:2021mxx,Nagle:2018nvi,Cunqueiro:2021wls,Altmann:2024icx,Grosse-Oetringhaus:2024bwr}. However, to date, there has been no measurable energy loss signal in these small systems. Understanding the system size dependence of these signals will help to answer whether and when the QGP is created in hadronic collisions.  

A promising new direction for studying QGP in small systems is the collision of light ions, e.g., oxygen-oxygen (OO)~\cite{Brewer:2021kiv} or neon-neon (NeNe)~\cite{Bally:2022vgo}. A short run of OO collisions at $\sqrt{s_\mathrm{NN}}=6.8\,\text{TeV}$ at LHC is planned in 2025~\cite{Citron:2018lsq,Bruce:2021hjk}, while the STAR experiment at RHIC has already taken data at $\sqrt{s_\mathrm{NN}}=200\,\text{GeV}$ in 2021~\cite{Liu:2022jtl,Huang:2023viw}.  Minimum bias OO collisions contain, on average, around ten participant nucleons, which is equivalent to the most peripheral lead-lead collisions~\cite{Huss:2020dwe}. As quenching scales with the system size~\cite{Mehtar-Tani:2021fud,Takacs:2021bpv,Mehtar-Tani:2024jtd,Shi:2022rja}, a small energy loss signal is expected in light ion collisions~\cite{Katz:2019qwv,Huss:2020whe,Zakharov:2021uza,Xie:2022fak,Ke:2022gkq,Vitev:2023nti,Ogrodnik:2024qug}. The successful discovery of quenching in small systems crucially depends on understanding the theory uncertainties of energy loss observables which is the main focus of this work.

In this work, we study in detail the jet nuclear modification factor $R_{\rm AA}$ and the jet-, and hadron-triggered semi-inclusive nuclear modification factors $I_\mathrm{AA}$ in OO collisions. We recall that previous measurements of inclusive~\cite{CMS:2016svx,ATLAS:2014cpa,ALICE:2021wct,ALICE:2023ama} and semi-inclusive~\cite{ALICE:2015mdb,ATLAS:2022iyq,ALICE:2017svf} nuclear modification factors have been used to set the upper limits on the energy loss in $p$Pb collisions.  Considering both LHC and RHIC kinematics, we calculate nuclear modification factors using next-to-leading-order (NLO) matrix elements with NLO nuclear parton distributions (nPDF) and NLO-matched parton showers with hadronization. We present systematic uncertainties in nPDF and scale components as well as sensitivity to parton showers and hadronization models. By neglecting any medium modification of final state interactions, we refer to this as the {\it no-quenching} baseline. 
Our finding shows that jet-triggered observables are much more robust than their hadronic counterpart. Furthermore, we present a kinematic window in which nPDF and scale uncertainties cancel dramatically in jet-triggered $I_{\rm AA}$ while showing little sensitivity to parton shower and hadronization models, addressing a major limiting factor for energy loss discovery in small systems.

The paper is organized as follows. In \cref{sec:RAA}, we present baseline computations of jet nuclear modification factor $R_\mathrm{AA}$ for OO collisions at $\sqrt{s_\mathrm{NN}} = 6.8\,\text{TeV}$ and $\sqrt{s_\mathrm{NN}} = 200\,\text{GeV}$. We compare nPDF uncertainties from different nPDF extractions and comment on scale, shower, and hadronization uncertainties. In \cref{sec:IAAjet}, we present the semi-inclusive jet-triggered nuclear modification factor and compute the baseline for the same center-of-mass energies. In particular, we analyze the (non)cancellation of nPDF uncertainties in different kinematic regions. We also study scale, parton shower, and hadronization uncertainties. In \cref{sec:IAAhadron}, we present the results for hadron-triggered $I_\mathrm{AA}$. We conclude in \cref{sec:conclusions}. A detailed description of event generator settings is included in \cref{app:egsettings}. \cref{app:absspectra} contains comparisons of computed absolute spectra for inclusive jet and jet-triggered semi-inclusive hadron cross sections with experimental data. A brief discussion of projected statistical uncertainties of the presented observables is given in \cref{app:statunc}.

\section{Jet nuclear modification factor $R_\mathrm{AA}^j$}\label{sec:RAA}

A standard way of measuring the modification of high momentum probes, e.g., jets or high $p_T$ hadrons, in nuclear collisions, is the (jet) nuclear modification factor
\begin{equation}\label{eq:RAA}
    R_\mathrm{AA}^{j} (p_{T},y) = \frac{1}{A^{2}} \frac{\dd\sigma^{j}_\mathrm{AA}/\dd p_{T}\,\dd y}{\dd\sigma^{j}_{pp}/\dd p_{T}\,\dd y} \, \;.
\end{equation}
It relates the inclusive jet cross section $\dd\sigma^{j}_\mathrm{AA}/\dd p_{T}\, \dd y$, differential in the transverse jet momentum $p_{T}$ and rapidity $y$, of a minimum bias AA collision with the one of a $pp$ collision $\dd\sigma^{j}_{pp}/\dd p_{T}\, \dd y$, normalized by the square of the nucleon number $\mathrm{A}^2$.

The inclusive jet cross section can be computed order by order using collinear factorization~\cite{Ellis:318585}
\begin{equation}
    \label{eq:inclusive_xsec}
    \sigma_{{\rm AA}\to j}=\sum_{i_1,i_2}\int \dd x_1\,\dd x_2\,f^{\rm A}_{i_1}\big(x_1,\mu_F^2\big)\,f^{\rm A}_{i_2}\big(x_2,\mu_F^2\big)\,\hat\sigma_{i_1i_2\to j}\big(\mu_R^2\big)\,,
\end{equation}
where $f^A_i$ denotes parton distributions, and $\hat\sigma$ is the partonic scattering amplitude of the $i_1i_2\to j$ process. The sum runs over all parton species, and $x_{1,2} = p^{\parallel}_{1,2}/P^{\parallel}_\mathrm{A}$ are fractions of the nucleus longitudinal momentum carried by the given parton species. The factorization scale $\mu_{F}$ marks the separation of short range partonic interactions and non-perturbative initial state effects, captured by the parton distribution functions. Finally, the coupling constant in $\hat\sigma$ is evaluated at the renormalization scale $\mu_{R}$. Additional power corrections can also appear beyond \cref{eq:inclusive_xsec}, including, for example, hadronization effects. 

We evaluate \cref{eq:inclusive_xsec} with the dijet production matrix element at next-to-leading-order (NLO) using \texttt{MadGraph5\_aMC@NLO}-\texttt{v2.9.18}~\cite{Alwall:2014hca} combined with NLO-matched \texttt{Pythia8.306} \cite{Sjostrand:2014zea,Bierlich:2022pfr} to receive showered and hadronized spectra. A detailed description of the event generator settings can be found in \cref{app:egsettings}. The factorization and renormalization scales were chosen to equal half the sum of the outgoing momenta $\mu_{F,R} =  \frac{1}{2} \sum_i p_{T,i}$. After showering, we achieve NLO+leading log accuracy.
We will use nPDF parametrizations implemented in the \texttt{LHAPDF6}~\cite{Buckley:2014ana} framework. Jets were clustered on all final state particles, using the anti-$k_{T}$ algorithm in \texttt{fastjet}~\cite{Cacciari:2008gp,Cacciari:2011ma} with $R = 0.4$ and $\vert y_\mathrm{jet} \vert < 3$. At NLO accuracy, the inclusive jet spectra reach excellent agreement with experimental data from $pp$ collision, see in~\cref{app:absspectra}.

Since the first baseline calculations~\cite{Huss:2020dwe,Brewer:2021tyv}, several nPDF groups have updated their global fits with new data from LHC. In this section, we provide an updated baseline for the jet nuclear modification factor for OO collisions at the expected $6.8$ TeV and 200 GeV center-of-mass energy, complementing the previous studies~\cite{Huss:2020dwe,Brewer:2021tyv,Belmont:2023fau}. We also study uncertainties of matrix elements, parton shower, and hadronization. For predictions of the expected energy loss signal in OO collisions, see Refs.~\cite{Huss:2020dwe,Zakharov:2021uza,Ke:2022gkq}.

\begin{figure}
    \centering
    \includegraphics[width=0.49\textwidth]{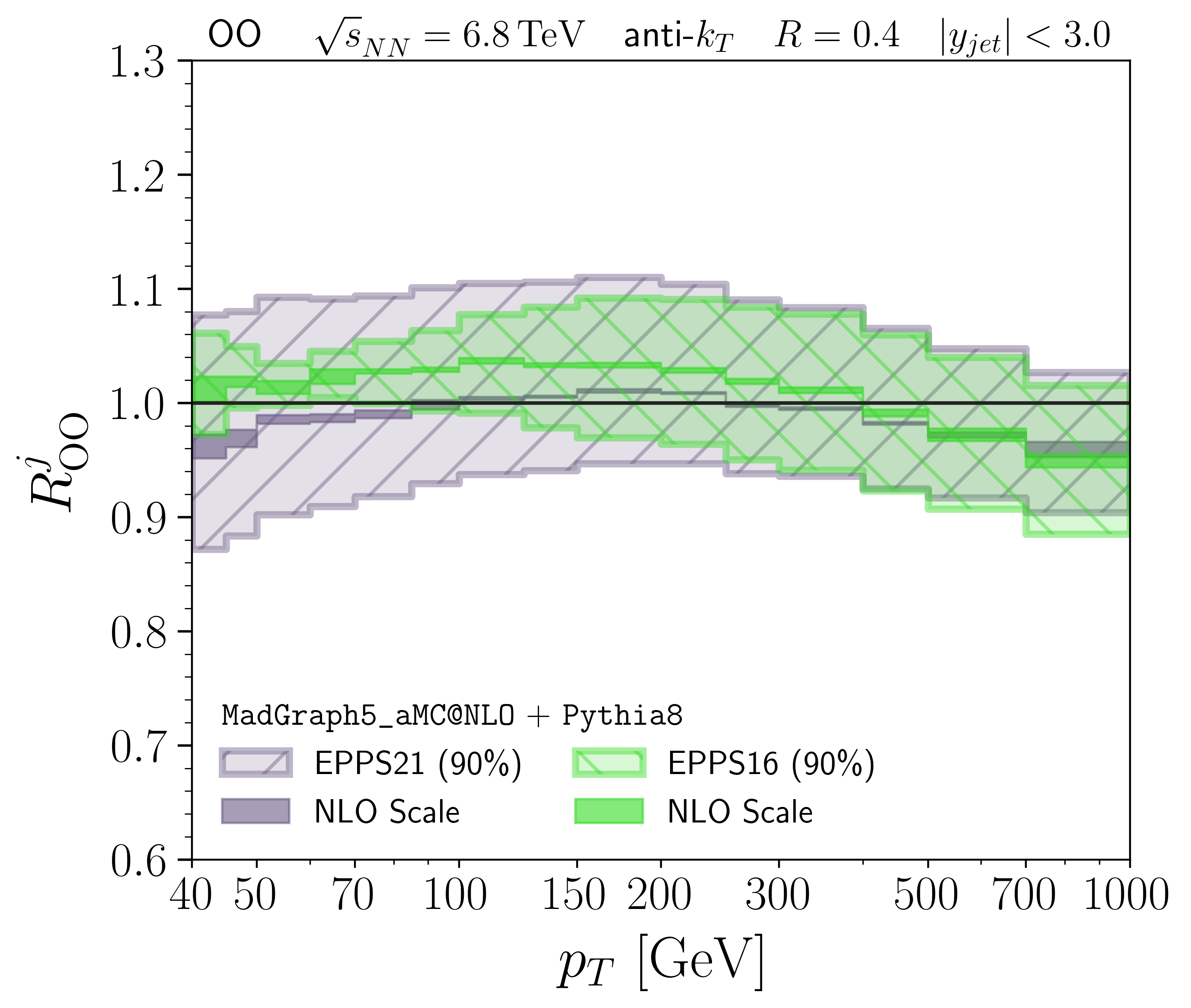}
    \includegraphics[width=0.49\textwidth]{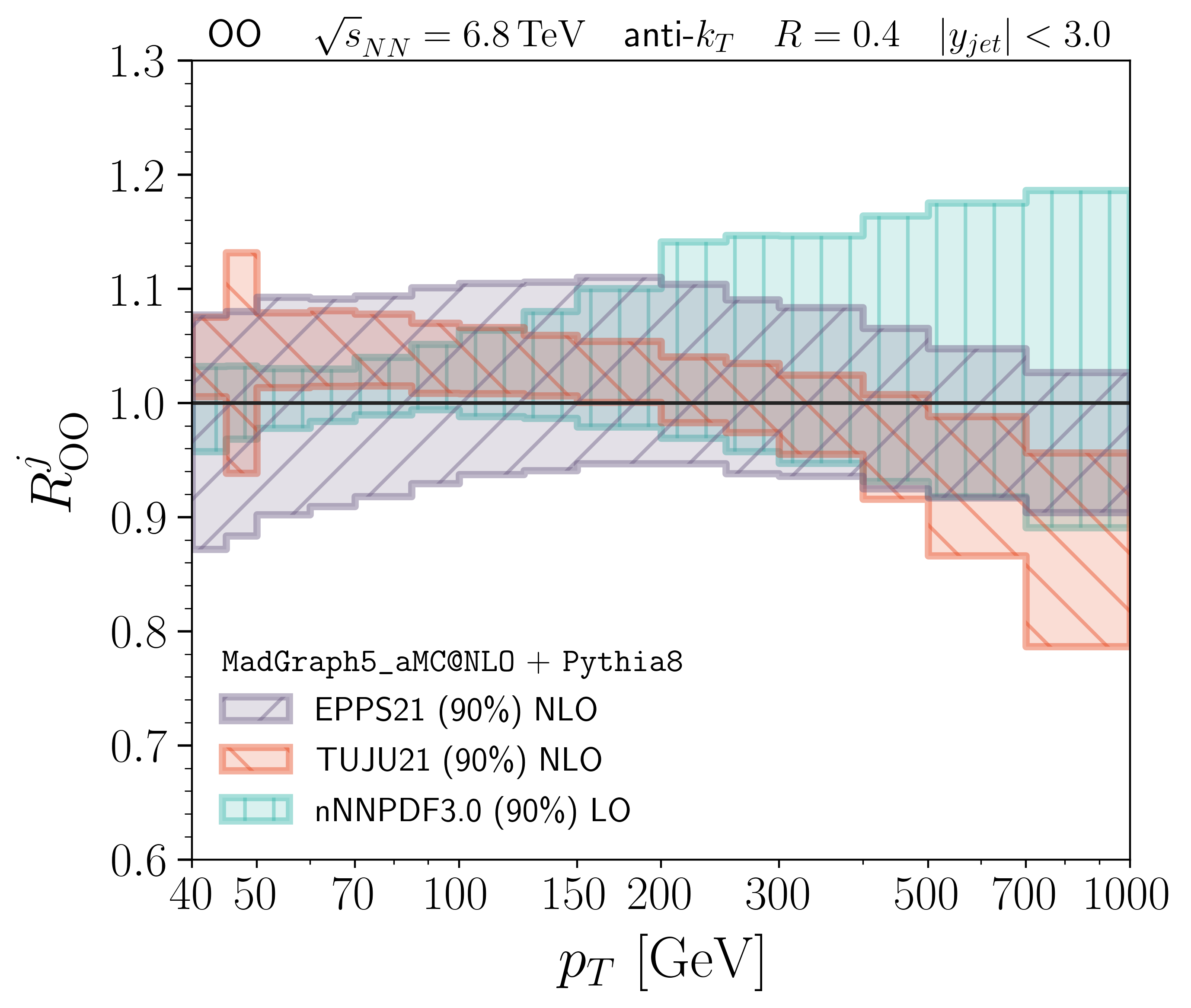}
    \caption{Predictions for the no-quenching baseline for jet nuclear modification factor in minimum bias oxygen-oxygen collisions at $\sqrt{s_\mathrm{NN}} = 6.8\,\text{TeV}$. Solid bands show 9-point scale variation and hatched bands indicate 90\% confidence interval of nPDF uncertainties. \textit{Left}: Comparison of EPPS16 and EPPS21. \textit{Right}: Comparison of EPPS21, TUJU21 and nNNPDF3.0.}\label{fig:RAA_6800}
\end{figure} 

The left panel of \cref{fig:RAA_6800} shows $R_\mathrm{OO}^{j}$ as a function of jet $p_T$ and compares results obtained with EPPS21~\cite{Eskola:2021nhw} to the older EPPS16~\cite{Eskola:2016oht} nPDF set. Correspondingly the $pp$ reference cross section is calculated with CT18ANLO~\cite{Hou:2019efy} and CT14NLO~\cite{Dulat:2015mca} free proton PDFs. The range of experimental data to constrain nPDFs in EPPS21 was extended to also contain recent DIS data, W$^{\pm}$, dijet and D$^{0}$ production. No final state interactions are included in our calculation, nevertheless the central value of $R_{\rm OO}\neq 1$ because of the nPDFs. The nPDF uncertainty of the two parametrizations differ especially at low $p_T$, where EPPS21 is four times more uncertain.
Larger uncertainties could originate from the updated $A$-dependence of the parametrization. The functional form in EPPS21 includes additional parameters (24 instead of 20), making the fit more flexible.
The measurement of dijet production in proton-oxygen collisions could be used to further constrain the nPDF uncertainties for light ions~\cite{Paakkinen:2021jjp}.
Uncertainties coming from scale variations of $\mu_F$ and $\mu_R$ are negligible compared to the nPDF uncertainties. 

The right panel of \cref{fig:RAA_6800} compares EPPS21, TUJU21~\cite{Helenius:2021tof} and nNNPDF3.0~\cite{AbdulKhalek:2022fyi} nPDFs. 
We observed occasional numerical instabilities in evaluating NLO uncertainties, leading to a sudden increase, as exemplified in the 45-50 GeV bin for TUJU21. Consequently, for nNNPDF3.0 we only show results obtained with matrix elements at leading-order (LO) combined with the NLO nPDF. Comparisons between LO and NLO matrix elements combined with the NLO nPDF results in almost identical $R_{\rm AA}$ (and $I_{\rm AA}$) nPDF uncertainty bands, see \cref{app:NLOvsLO}. While central values are quite different, TUJU21 and nNNPDF3.0 predictions are mostly contained within the EPPS21 uncertainty band below $p_T = 200\,\mathrm{GeV}$. Above, TUJU21 and nNNPDF3.0 uncertainties become larger but all predictions are still consistent with each other.
We note that TUJU21 extractions use only 16 free parameters in contrast to 24 of EPPS21. Therefore, smaller uncertainties could result from more restrictive parametrization. On the other hand, nNNPDF3.0 uses a neural network-based approach that can incorporate a much larger number of parameters and fewer assumptions on the functional form of the nPDFs. Therefore nNNPDF3.0 uncertainties can grow rapidly in kinematic regions not constrained by experimental data.

\begin{figure}
    \centering
    \includegraphics[width=0.47\textwidth]{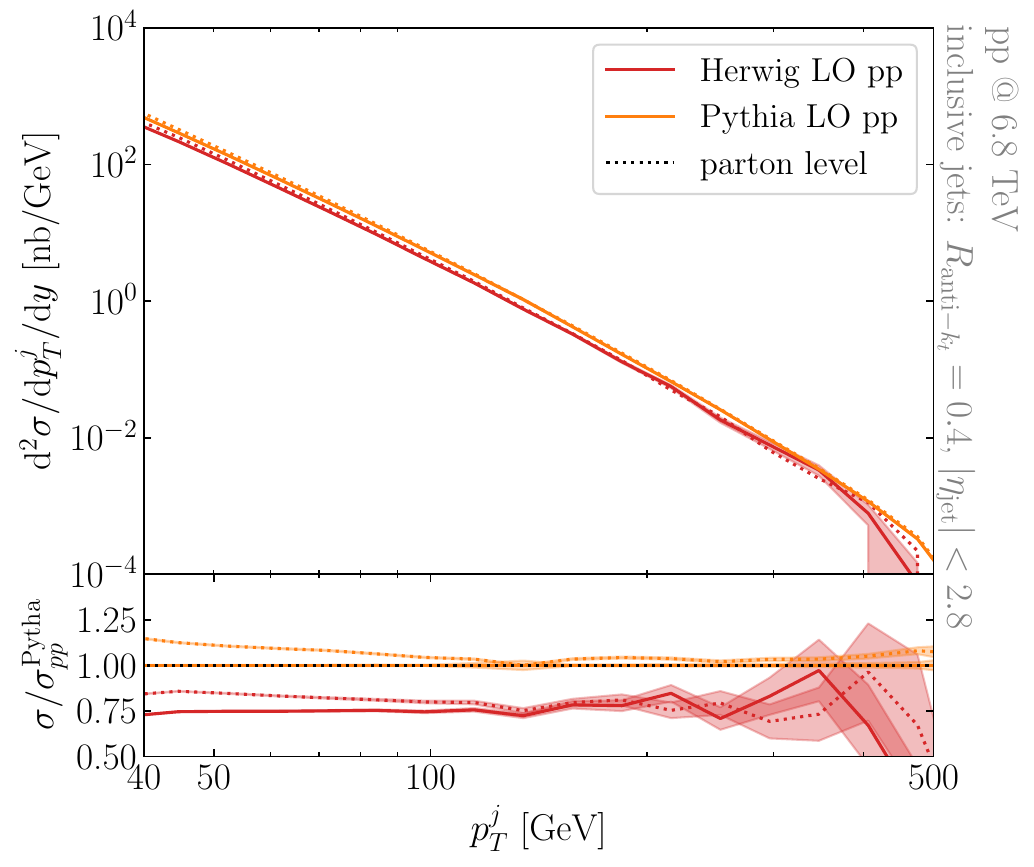}
    \includegraphics[width=0.52\textwidth]{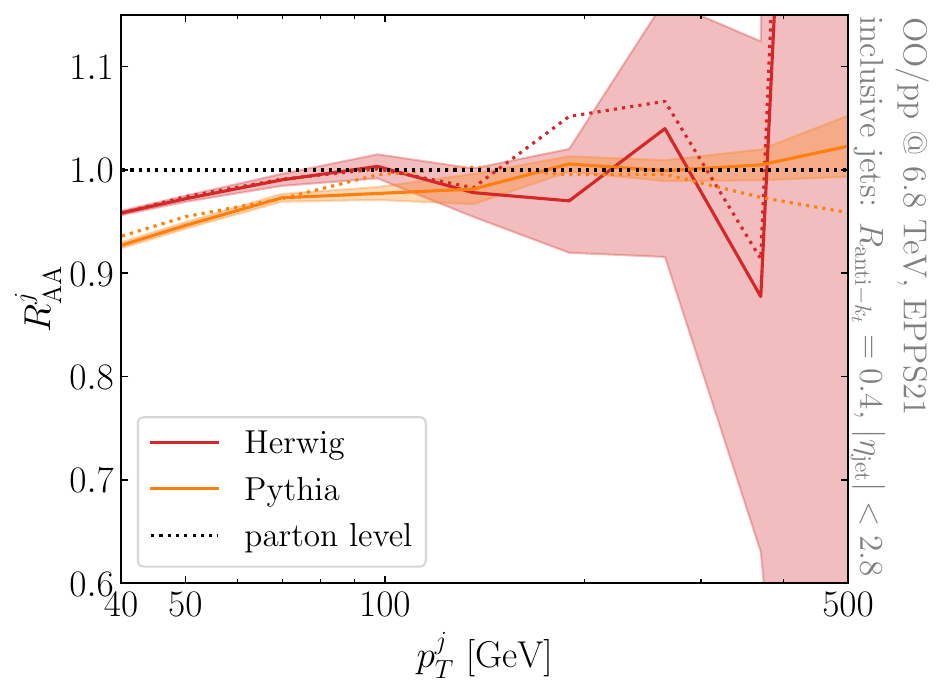}
    \caption{The jet spectrum and the nuclear modification factor in minimum bias oxygen-oxygen collisions at $\sqrt{s_\mathrm{NN}} = 6.8\,\text{TeV}$ with Pythia and Herwig. Jets are reconstructed on all hadrons (full lines), or on partons (dotted lines). Solid bands show statistical uncertainty.}
    \label{fig:spectra_Pythia_Herwig}
\end{figure} 

To estimate shower and hadronization uncertainties, we show the inclusive jet spectrum and the $R_{\rm AA}^j$ in $pp$ and OO collisions in the \texttt{Pythia8.312} and \texttt{Herwig7.3.0}~\cite{Bahr:2008pv,Bellm:2015jjp} event generators in \cref{fig:spectra_Pythia_Herwig}. We use LO matrix elements, CT18ANLO PDFs, and NLO EPPS21 nPDFs (LO nPDFs are not available). While the shape of the spectrum is similar, the magnitude differs in the two event generators. This difference is significantly reduced in $R_{\rm AA}$ resulting in a a few percent difference. Hadronic jet spectra differ slightly from the parton level distributions (dotted lines). It mostly affects the jet spectrum at lower momenta, while hadronization corrections are negligible in $R_{\rm AA}$. This robustness against hadronization effects is due to the infrared and collinear safe properties of jets.

\begin{figure} 
    \centering
    \includegraphics[width=0.49\textwidth]{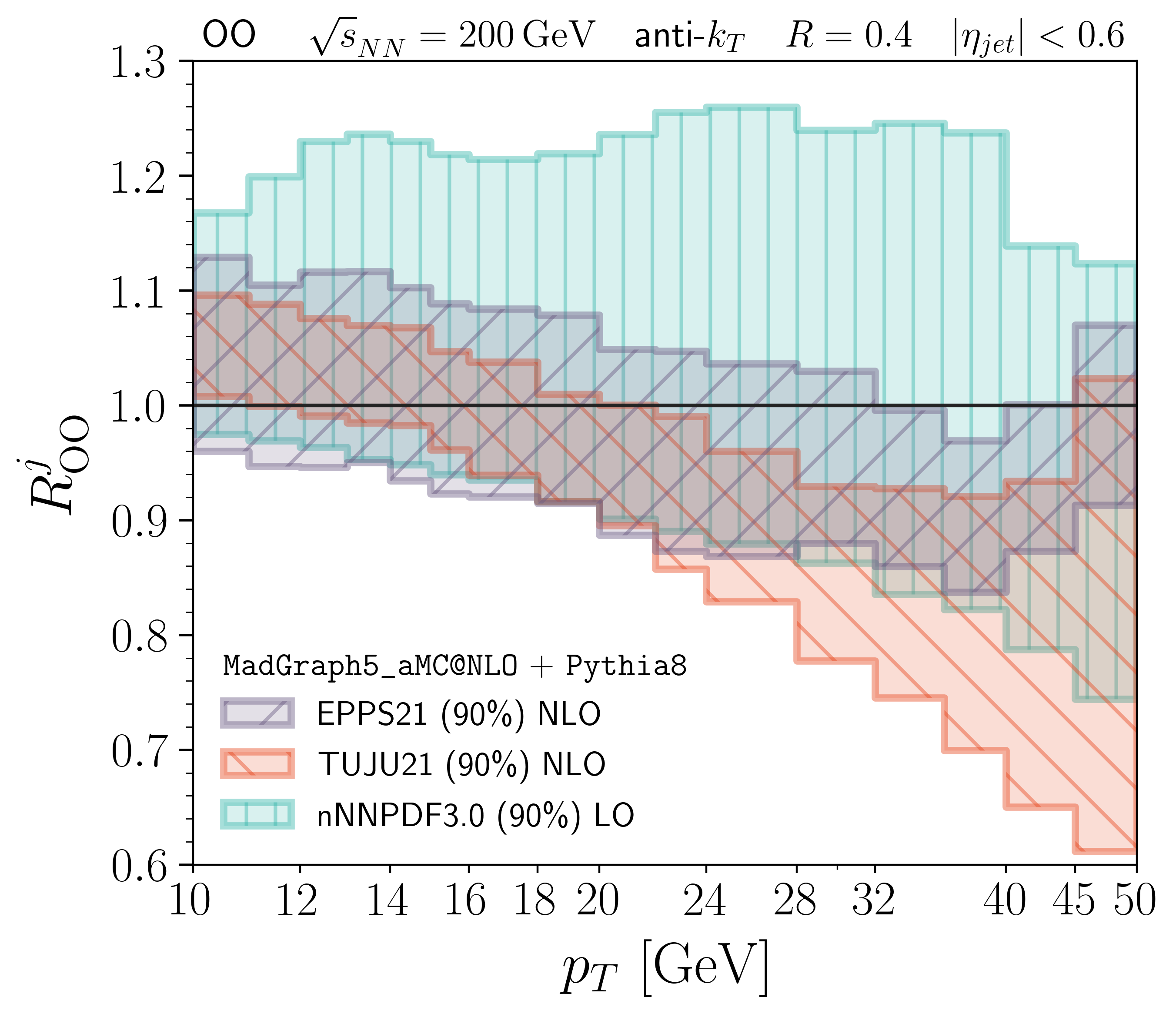}
    \caption{NLO predictions for the no-quenching baseline for the charged jet nuclear modification factor in minimum bias OO collisions at $\sqrt{s_\mathrm{NN}} = 200\,\text{GeV}$. Hatched bands indicate 90\% confidence interval of nPDF uncertainties.} \label{fig:RAA_200}
\end{figure} 

Baseline calculations for charged jet $R_\mathrm{OO}^j$ at $\sqrt{s_\mathrm{NN}} = 200\,\text{GeV}$ are shown in \cref{fig:RAA_200}. In this case, the clustering was performed on charged particles, requiring $\vert \eta_\mathrm{jet} \vert < 0.6$. Deviation of $R_\mathrm{OO}^j$ from unity is clearly present due to the nPDFs. As in \cref{fig:RAA_6800}, TUJU21 tends to have smaller uncertainties than EPPS21, except for large momentum. nNNPDF3.0 errors are large across all shown $p_T$.
The central values differ among nPDF sets but still agree within errorbands. For discussion of the seen negative trend in $R_\mathrm{OO}^j$ see \cref{ssec:pnpunc}.

\section{Jet-triggered hadron nuclear modification factor $I_\mathrm{AA}^h$}
\label{sec:IAAjet}

\subsection{No-quenching baseline results}
\label{sec:IAA_nPDF_uncertainty}

The semi-inclusive nuclear modification factor $I_\mathrm{AA}$ was proposed as a self-normalizing observable~\cite{ALICE:2015mdb}. For example, the jet-triggered hadron spectrum $I_{\rm AA}$ is 
\begin{equation}
    I_\mathrm{AA}^{h} = \frac{{1}/{\sigma^{j}_\mathrm{AA}}~ {\dd\sigma^{h+j}_\mathrm{AA}}/{\dd p_{T}}}{{1}/{\sigma^{j}_{pp}}~ {\dd\sigma^{h+j}_{pp}}/{\dd p_{T}}}\;,
\end{equation}
where the trigger jet yield $\sigma^{j}$ is used as normalization. In particular, measurements of semi-inclusive nuclear modification factor have been previously used to place upper limits to energy loss in $p$Pb collisions~\cite{CMS:2016svx,ATLAS:2014cpa,ALICE:2021wct,ALICE:2023ama}.

We evaluate $I_{\rm AA}$ using the same NLO+LL setup as earlier with details described in \cref{app:egsettings}. The main difference lies in the more differential $h+j$ final state~\cite{deFlorian:2009fw}.
We follow the conventions of Ref.~\cite{ATLAS:2022iyq}, and we calculate the jet-triggered hadron spectrum in OO collisions at $\sqrt{s_\mathrm{NN}} = 6.8\,\text{TeV}$. Jet triggers are clustered from all particles using the anti-$k_T$ algorithm with $R = 0.4$ and satisfy the momentum cut $p_T^j>p_{T,\text{min}}^j$ and $|\eta_\mathrm{jet}|<2.8$. Then charged hadrons are selected with $|\eta|<2.035$. For each jet that satisfies the trigger, we compute the momentum differential cross section of all charged hadrons near- and away-side of the jet. The total away-side yield reads
\begin{equation}\label{eq:YAAjetaway}
    Y_\mathrm{AA,\,away}^{h}(p_{T}) ~=~ \frac{1}{\sigma^{j}_\mathrm{AA}} \frac{\dd\sigma^{h+j}_\mathrm{AA}}{\dd p_{T}} ~ \Bigg\vert_{\, p_{T}^{j} > p_{T,\mathrm{min}}^{j}\, \;, ~ \Delta\Phi > \frac{7\pi}{8} } \, \;,
\end{equation}
where $\sigma^{j}_\mathrm{AA}$ is the inclusive jet cross section for $p_{T}^{j} > p_{T,\mathrm{min}}^{j}$ and $\dd\sigma^{h+j}_\mathrm{AA}/\dd p_{T}$ is the cross section of finding a correlated charged hadron with transverse momentum $p_{T}$. The $\Delta \Phi$ is the minimal \textit{azimuthal} angle in the transverse plane between the jet axis and a hadron,
\begin{equation}
    \label{eq:deltaphi}
    \Delta \Phi(\Phi_{h},\!\Phi_{j}) = \left\{
    \begin{array}{c}
         \begin{aligned}
             &\vert \Phi_{h}\! - \! \Phi_{j} \! - \! 2\pi  \vert \;, ~~ \Phi_{h} \! - \! \Phi_{j} > \! \pi \, \, \\[2pt]
             &~~ \vert \Phi_{h} \! - \! \Phi_{j} \vert \;, \, -\pi < \! \Phi_{h} \! - \! \Phi_{j} < \! \pi \\[2pt]
             &\vert \Phi_{h} \! - \! \Phi_{j} \! + \! 2\pi  \vert \;, ~ \Phi_{h}\! - \! \Phi_{j} < \! -\pi 
         \end{aligned}
    \end{array}
    \right\}.
\end{equation}
A hadron is said to be opposite (or away-side) if $\Delta \Phi > 7\pi/8$ and near-side for $\Delta \Phi < \pi/8$. Hence, the near-side yield is given by
\begin{equation}
    \label{eq:YAAjetnear}
    Y_\mathrm{AA,\,near}^{h}(p_{T}) ~=~ \frac{1}{\sigma^{j}_\mathrm{AA}} \frac{d\sigma^{h+j}_\mathrm{AA}}{dp_{T}} ~ \Bigg\vert_{\, p_{T}^{j} > p_{T,\mathrm{min}}^{j}\, \;, ~ \Delta\Phi < \frac{\pi}{8} } 
\end{equation}
and tells how charged hadrons are distributed around a given jet.
The semi-inclusive nuclear modification factor is given by the ratio 
\begin{equation}\label{eq:IAA}
    I_\mathrm{AA}^{h}(p_T) = \frac{Y_\mathrm{AA}^{h}(p_T)}{Y_{pp}^{h}(p_T)}\,.
\end{equation}

\begin{figure*}
    \centering
    \includegraphics[width=0.49\textwidth]{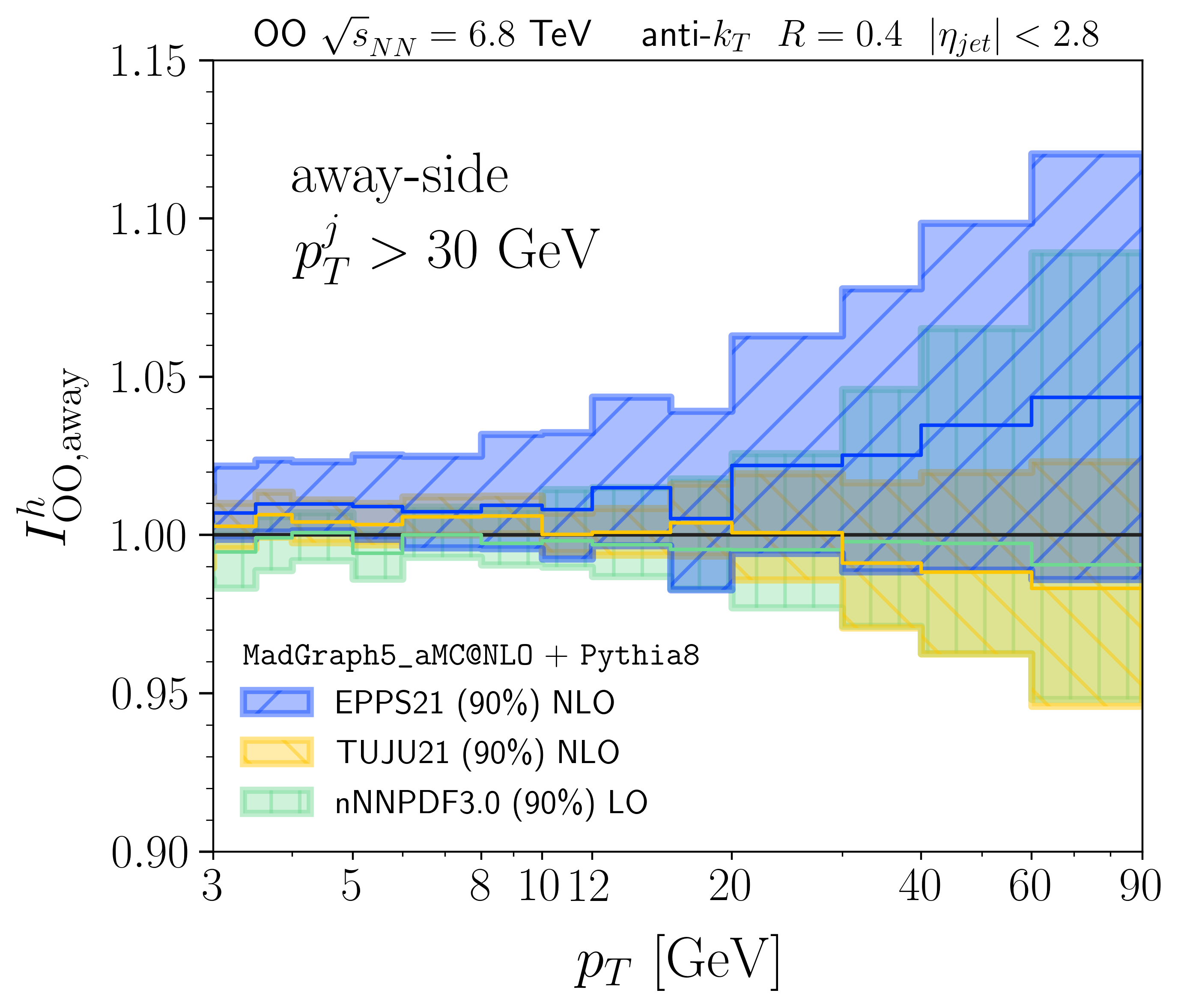}
    \includegraphics[width=0.49\textwidth]{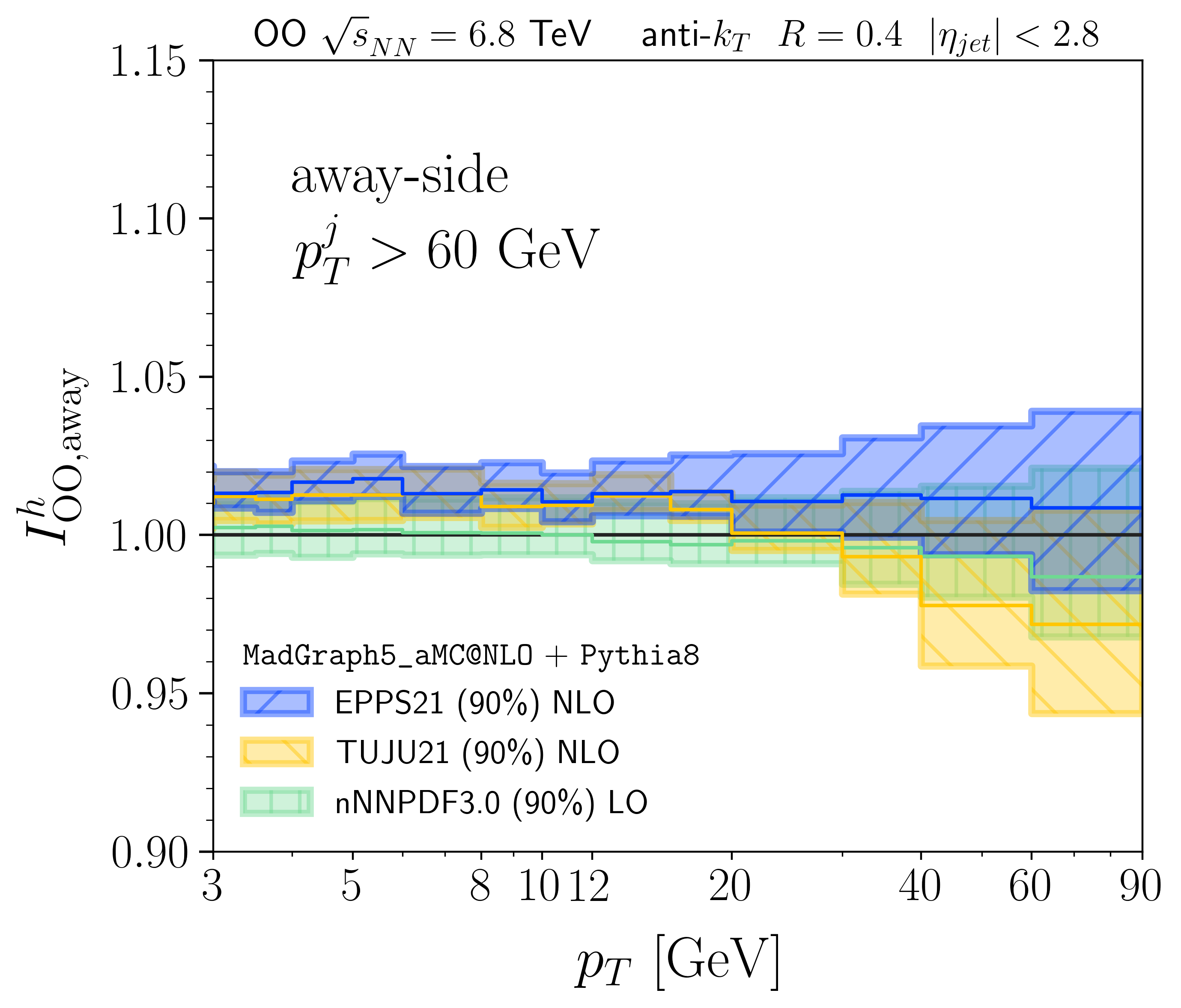}
    \includegraphics[width=0.49\textwidth]{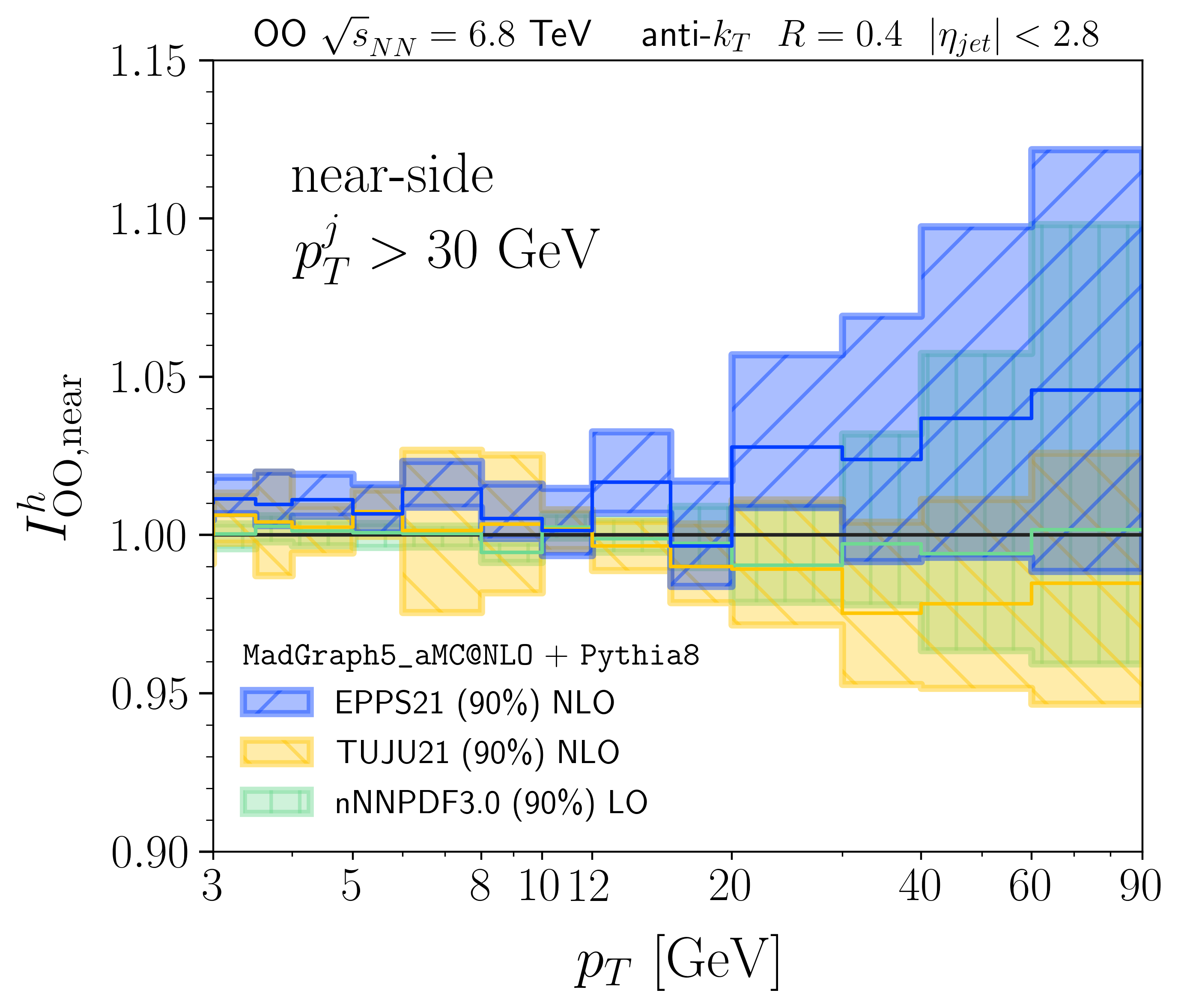}
    \includegraphics[width=0.49\textwidth]{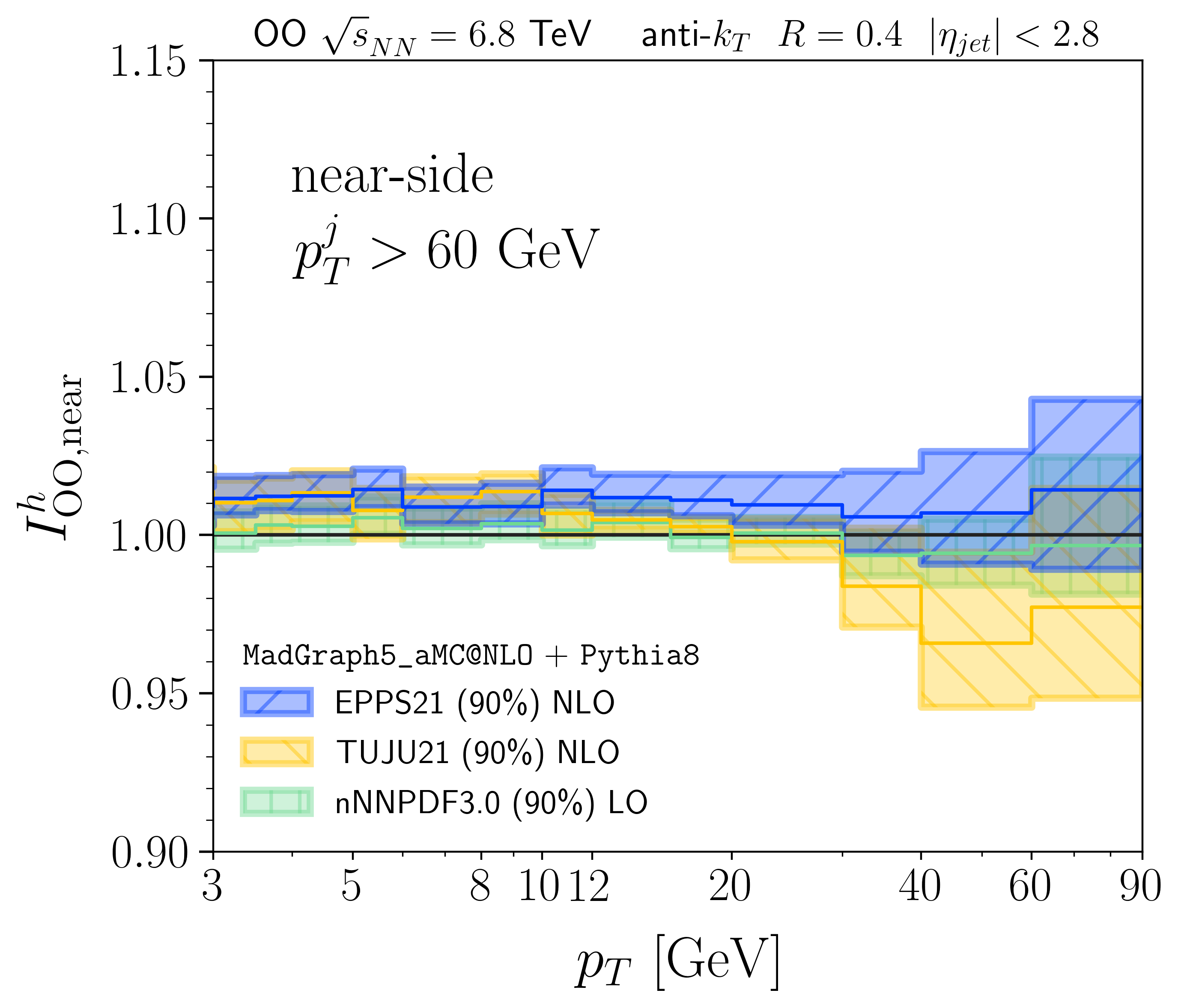}
    \caption{No-quenching baseline for jet-triggered hadron $I^{h}_\mathrm{AA}$ in OO collisions at $\sqrt{s_\mathrm{NN}} = 6.8\,\text{TeV}$. The {\it left} column shows results for jet trigger momentum $p_T^j>30\,\text{GeV}$, while the {\it right} column for $p_T^j>60\,\text{GeV}$. {\it Upper} and {\it lower} rows show away-, and near-side hadrons relative to the trigger. The solid lines correspond to the central nPDF values, whereas the bands show nPDF uncertainties at a $90\%$ confidence level.} 
    \label{fig:jet-hadronat6800}
\end{figure*} 

The four panels in \cref{fig:jet-hadronat6800} compare the no-quenching baseline of $I_\mathrm{OO}^h$ for near- and away-side yields obtained with different oxygen nPDFs and for different jet-momentum triggers at 6.8\,TeV center-of-mass energy\footnote{\label{note1}When comparing computed total yields $Y_\mathrm{AA}^h$ with measurements in $pp$ collisions in \cref{app:absspectra}, we find that the computations over predict the amount of produced hadrons but this difference in magnitude cancels after taking the ratio to construct the semi-inclusive modification factor $I_\mathrm{AA}^h$.}. The central values deviate from unity because of the nPDFs and they vary for different nPDF parametrizations at high $p_T$. EPPS21 results tend to become larger, while TUJU21 and nNNPDF3.0 are moderately smaller than unity. 
The nPDF uncertainties are smaller than those for jet $R_{\rm OO}$ in \cref{fig:RAA_6800}. In some kinematic region, they are ten times smaller, demonstrating the resilience of the double ratio structure. As $p_T$ becomes larger, nPDF uncertainties show significant growth. The characteristics of near- and away-side $I_\mathrm{OO}^{h}$ are similar (upper versus lower panels). 
Higher trigger jet momenta (right panels), significantly decrease nPDF uncertainties for shown $p_T$. In \cref{sec:cancellation} we give a detailed explanation for the trends seen in uncertainty cancellation and baseline deviation from unity.

\begin{figure*}
    \centering
    \includegraphics[width=0.49\textwidth]{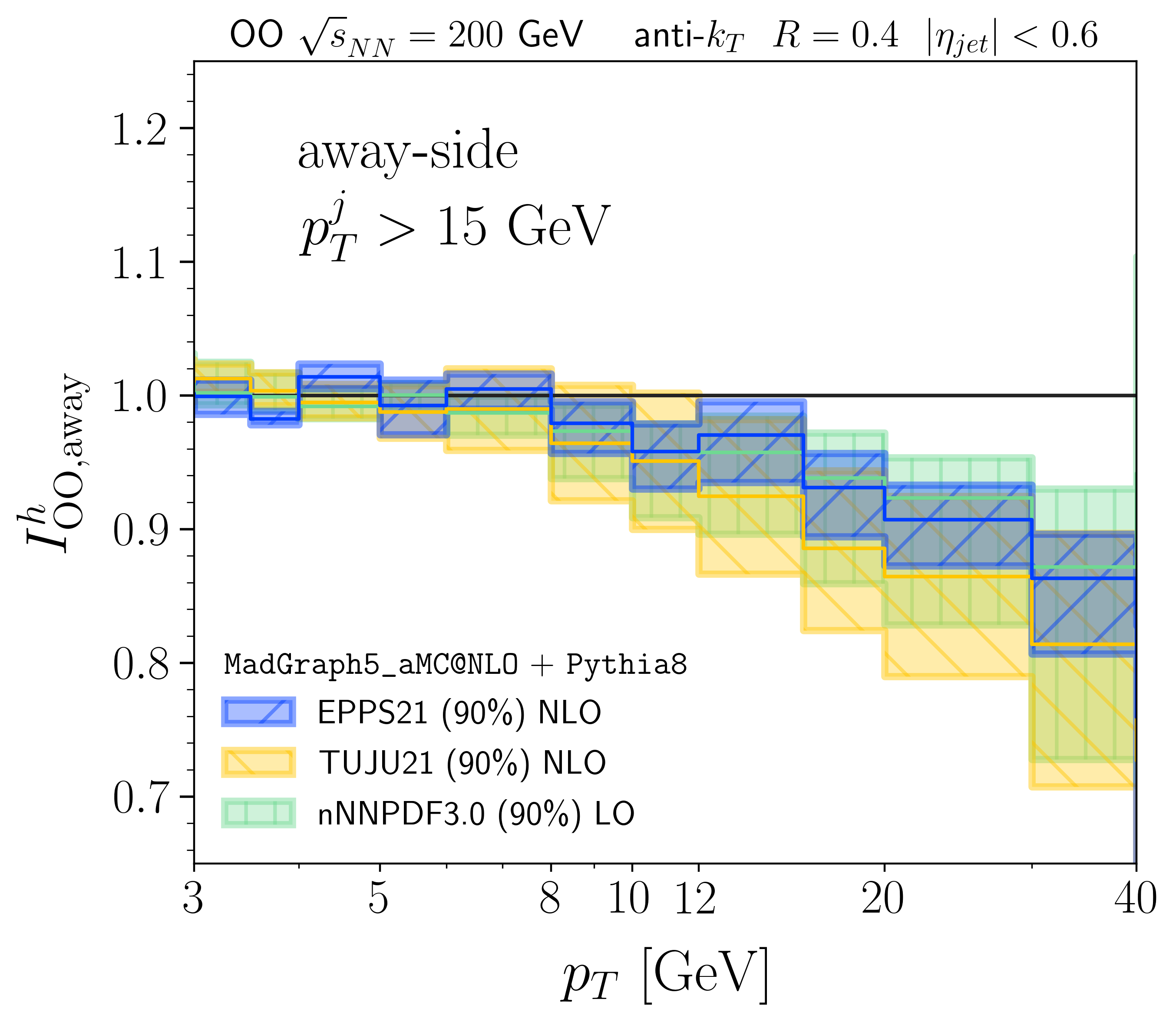}
    \includegraphics[width=0.49\textwidth]{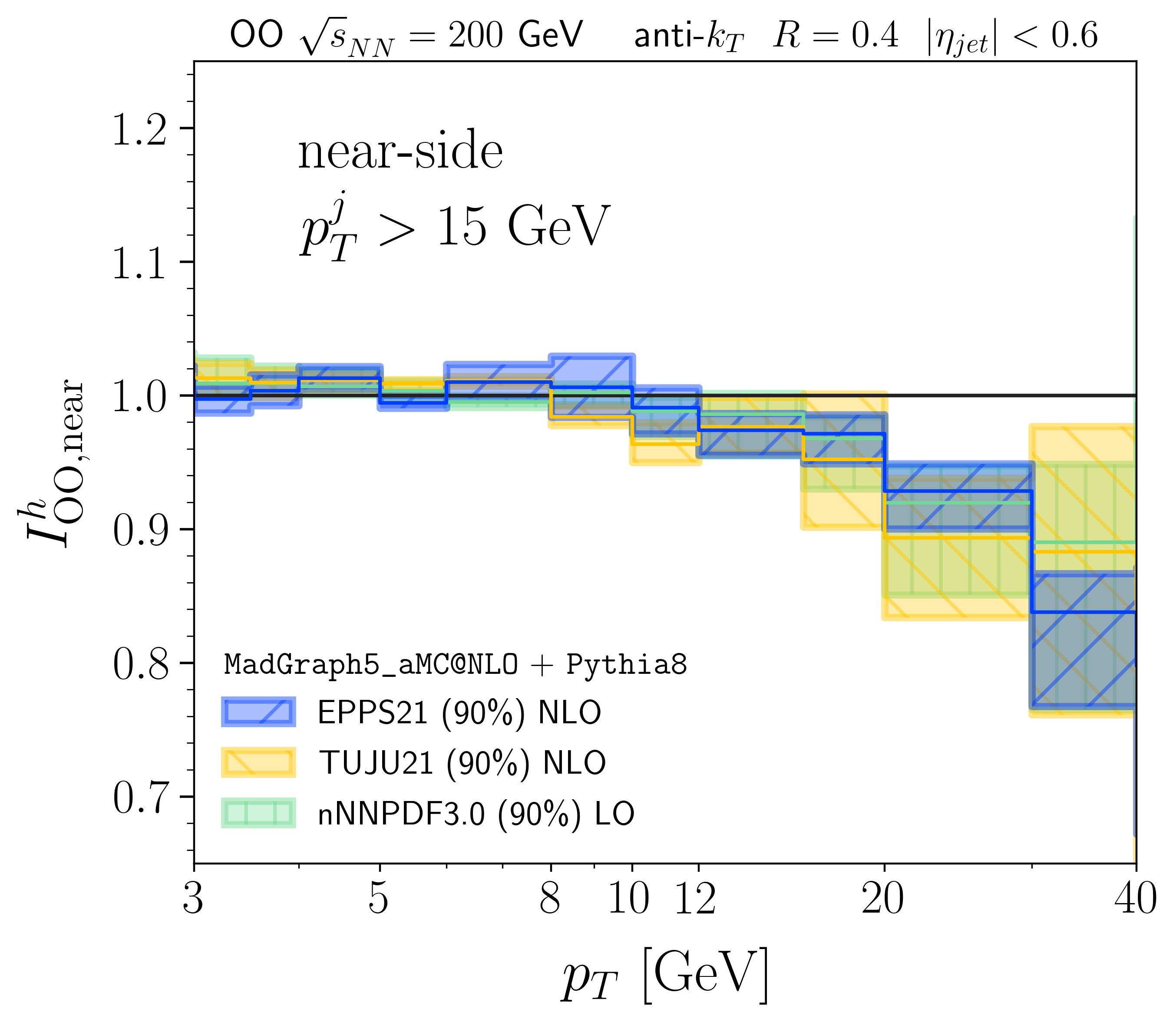}
    \caption{
    No-quenching baseline for jet-triggered hadron $I^{h}_\mathrm{AA}$ in OO collisions at $\sqrt{s_\mathrm{NN}} = 200\,\text{GeV}$ for jet trigger momentum $p_T^j>15\,\text{GeV}$. {\it Left and right} panels show away-, and near-side hadrons relative to the trigger.
    } 
    \label{fig:jet-hadronat200}
\end{figure*}

In \cref{fig:jet-hadronat200} we show analogous results at 200\,GeV center-of-mass energy. Here, jets were clustered on charged particles, and due to the lower energy, we consider a lower trigger cut-off of $p^j_{T,\mathrm{min}}=15\,\text{GeV}$. Jets must lie within rapidity range $|\eta_{\rm jet}|<0.6$ and charged hadrons within $|\eta|<1.0$. For higher momentum, the no-quenching baseline is significantly suppressed below unity because of the nPDF effects.

\subsection{Cancellation of nPDF uncertainty and deviation from unity}\label{sec:cancellation}

\Cref{fig:jet-hadronat6800,fig:jet-hadronat200} showed that nPDF uncertainties are strongly suppressed in particular kinematics, while the central value can deviate from unity. In this section, we explain these phenomena.

We first recall how nPDF uncertainties are actually computed. EPPS and TUJU collaborations use the \textit{Hessian} method \cite{Pumplin:2001ct} to compute the uncertainties. The central (best fit) nPDF $f^0$ is accompanied by an error set with twice as many members as the number of fitting parameters ($f_{k}^{\pm}$, $k=1,\dots,N_{\rm fit}$). Each member in the error set corresponds to variations along $N_\mathrm{fit}$ orthogonal eigenvectors in the fit parameter space that increase the chi-squared by a given tolerance.
The asymmetric uncertainty of some observable $X[f]$ is then computed according to 
\begin{equation}\label{eq:errorasym} 
    \begin{aligned}
        \Delta X^{+} ~&=~ \sqrt{\sum\limits_{k=1}^{N_{\rm fit}} \bigg[ \max \Big( X[f^{+}_{k}] - X[f^{0}], X[f^{-}_{k}] - X[f^{0}], 0 \Big) \bigg]^{2}}\,\;, \\
        \Delta X^{-} ~&=~ \sqrt{\sum\limits_{k=1}^{N_{\rm fit}} \bigg[ \min \Big( X[f^{+}_{k}] - X[f^{0}], X[f^{-}_{k}] - X[f^{0}], 0 \Big) \bigg]^{2}} \, \;.
    \end{aligned}
\end{equation}
EPPS and TUJU collaborations provide error sets that produce $90\%$ confidence intervals.
When computing uncertainties of $I_{\rm AA}$, we first compute $I_{\rm AA}$ for each member of nucleus nPDF and proton PDF sets. Note that the uncertainties in the reference proton PDF are propagated to nPDF sets. Therefore $I_{\rm AA}$ is computed using matching members of the proton and nucleus error sets and only then \cref{eq:errorasym} is applied. 

The neural network-based nNNPDF framework uses the \textit{replica} method~\cite{Forte:2002fg}.
The error set corresponds to the Monte Carlo sample of
the probability density of the fitted data.
The $X\%$ confidence interval of some observable $X[f]$ is computed by sorting the values of the error set $X[f_i]$ and removing symmetrically $(100 - X)\%$ members of the replicas with the highest and lowest values. The central value is the median value of the error set.

To gain a simple intuition in uncertainty cancellation, consider a more straightforward case where uncertainties can be computed from a variance over the members of the error set, i.e., $\Delta X\equiv \mathrm{var}(X[f])$. Then the uncertainty of the ratio $X/Y$ is given by
\begin{equation}
   \Delta\left(\frac{X}{Y}\right) =\frac{X}{Y} \sqrt{\left( \frac{\Delta X}{X}\right)^2+\left( \frac{\Delta Y}{Y}\right)^2-2\rho(X,Y) \frac{\Delta X}{X}\frac{\Delta Y}{Y} }\,,
\end{equation}
where $\rho$ is the Pearson correlation coefficient,
\begin{equation}
    \label{eq:pearson0}
    \rho(X,Y) ~=~ \frac{\mathrm{cov}(X,Y)}{\sqrt{\mathrm{var}(X)\,\mathrm{var}(Y)}} \, \;.
\end{equation} 
We see that the cancellation of uncertainties in the ratio $X/Y$ is only possible when the correlation is positive $\rho>0$. If $\rho<0$, we have the case when the uncertainties in the ratio even increase.

\begin{figure*}
    \centering
    \includegraphics[width=0.49\textwidth]{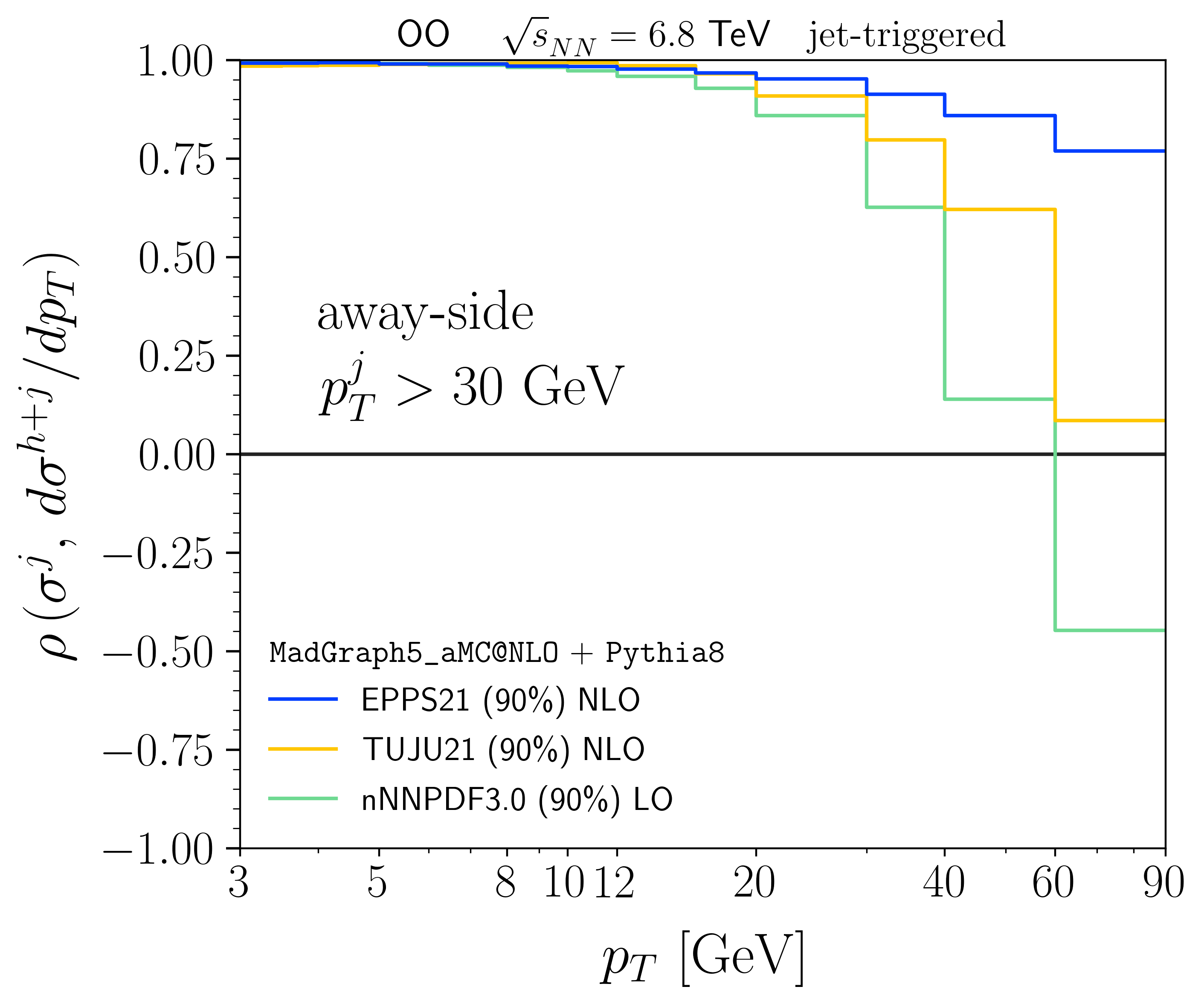}
    \includegraphics[width=0.49\textwidth]{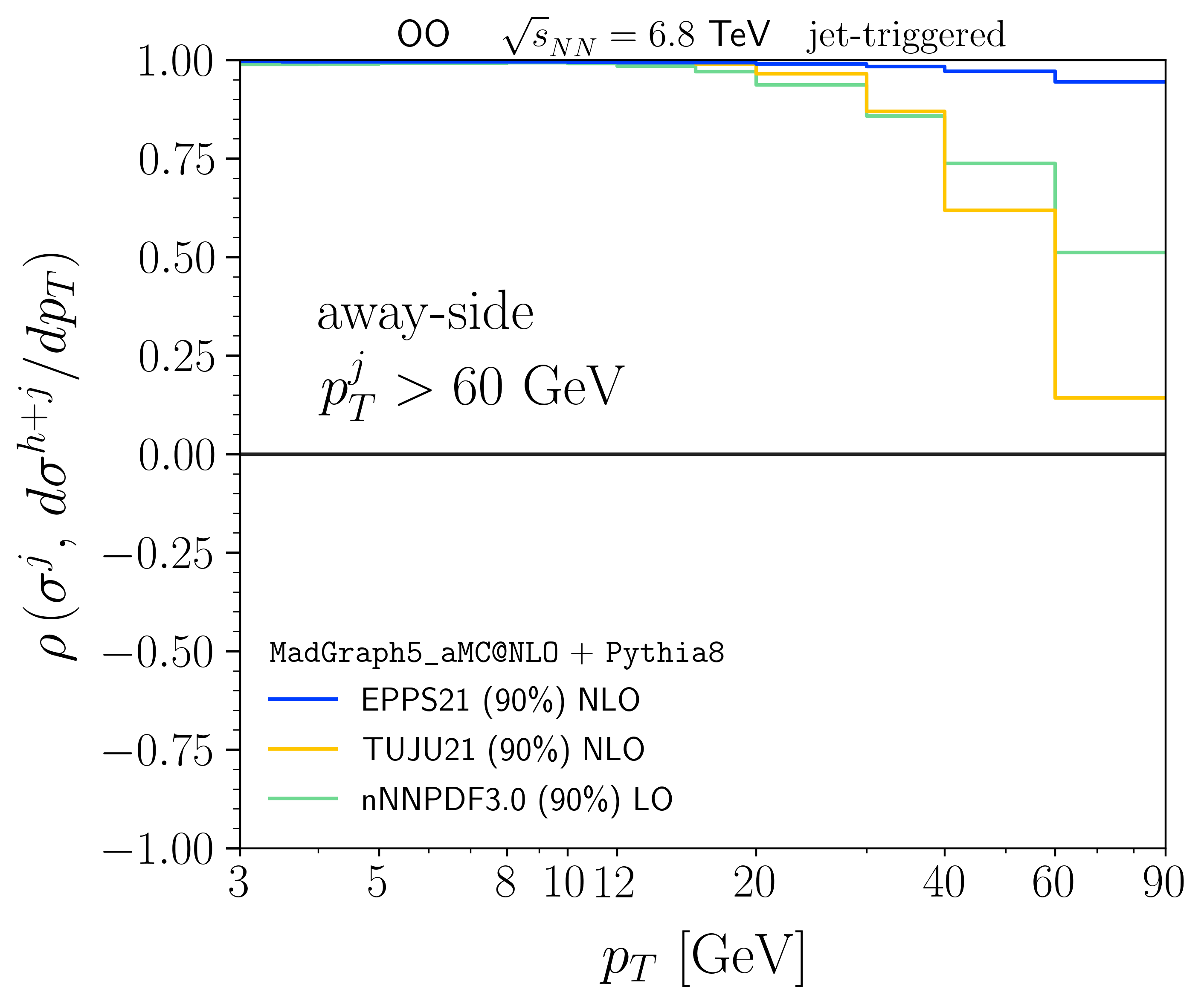}
    \caption{The Pearson correlation coefficient of $\sigma_\mathrm{OO}^{j}$ and $\dd\sigma^{h+j}_\mathrm{OO,away}/\dd p_{T}$ for $p_{T,\mathrm{min}}^{j} = 30\,\text{GeV}$ (left) and $p_{T,\mathrm{min}}^{j} = 60\,\text{GeV}$ (right). Differently colored curves represent different nPDF choices.} 
    \label{fig:pearsonjet}
\end{figure*}

Because $Y_{pp}$ does not depend on the nPDF, the uncertainty of $I_{\rm AA}$ is especially sensitive to the cancellation of nPDF uncertainties between $\sigma^{j}_{\rm AA}$ and $\sigma^{h+j}_{\rm AA}$. To study this, we compute the Pearson correlation,
\begin{equation}
    \label{eq:pearson}
    \rho(\sigma^{j}_{\rm AA},\sigma^{h+j}_{\rm AA}) ~=~ \frac{\mathrm{cov}(\sigma^{j}_{\rm AA},\sigma^{h+j}_{\rm AA})}{\sqrt{\mathrm{var}(\sigma^{j}_{\rm AA})\,\mathrm{var}(\sigma^{h+j}_{\rm AA})}} \, \;,
\end{equation} 
using the members of nPDF uncertainty sets\footnote{Variance of Hessian PDF error sets does not give the correct uncertainty, cf.~\cref{eq:errorasym}, therefore our results are only indicative. In addition, nPDF error sets also include uncertainties due to the proton reference, but this uncertainty cancels in the ratio between $Y_{\rm AA}$ and $Y_{pp}$ to a large degree.}. 
\Cref{fig:pearsonjet} shows the Pearson correlation for LHC kinematics (RHIC kinematics would conclude the same). The nPDF uncertainties are strongly correlated below 30 GeV, explaining the observed uncertainty cancellation in \cref{fig:jet-hadronat6800,fig:jet-hadronat200}. As $p_{T}$ approaches $p_{T,\mathrm{min}}^{j}$, the correlation decreases visibly increasing the nPDF uncertainty. Increasing $p^j_{T,\mathrm{min}}$ on the right panel of \cref{fig:pearsonjet}, we find large positive correlation over a wider range of hadronic transverse momenta, consistent with broad uncertainty cancellation visible in the right panels of \cref{fig:jet-hadronat6800}. 

\begin{figure*}
    \centering
    \includegraphics[width=\textwidth]{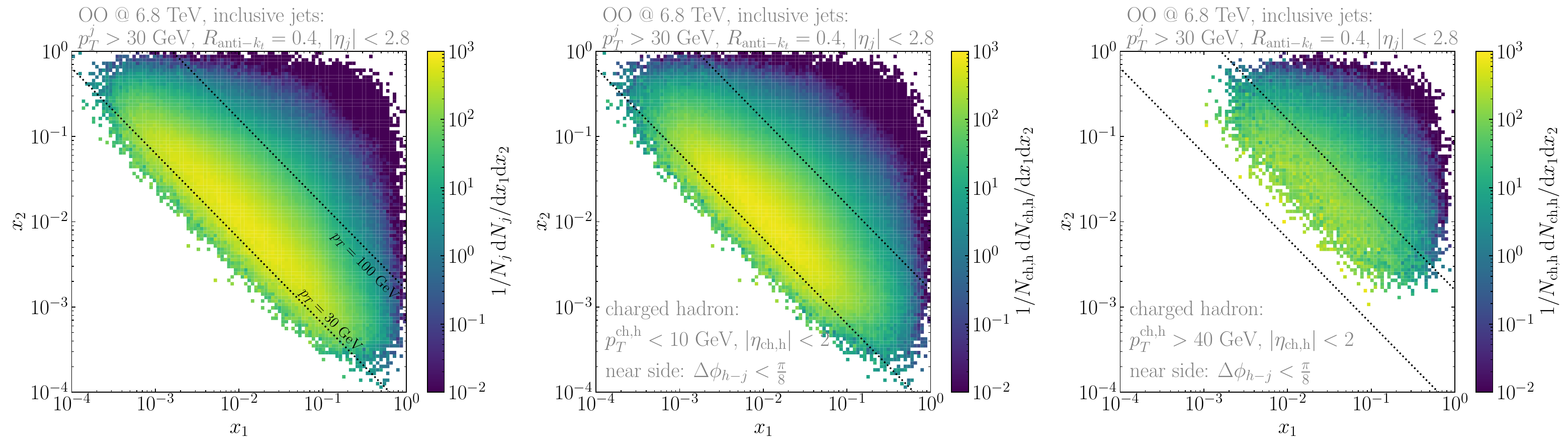}
    \caption{The probed $x_1$ and $x_2$ values in the PDFs in oxygen-oxygen collisions for different event selections, using the \texttt{Pythia8} event generator. \textit{Left:} Inclusive jet cross section with $p_T^j>30\,\text{GeV}$, \textit{Center:} Semi-inclusive cross section of low momentum hadrons $p_T<10\,\text{GeV}$, and \textit{Right:} High momentum hadrons $p_T>40\,\text{GeV}$.} 
    \label{fig:fullbjorkenx}
\end{figure*} 

In the following, we will use parton kinematics to explain the loss of uncertainty correlations between high momentum hadrons and trigger jet cross section. The nPDFs in the factorization formula \cref{eq:inclusive_xsec} are evaluated at Bjorken $x_{1,2}$, where at leading order we have~\cite{Ellis:318585}
\begin{equation}
    \label{eq:bjorkenx}
    x_1 x_2 ~=~ \frac{2p_T^2}{s_{\rm NN}} \Big(1 + \cosh(y_1 - y_2)\Big) \geq \frac{4p_T^2}{s_\mathrm{NN}} \, \;.
\end{equation}
Here, $p_T$ and $y_{1,2}$ denote the transverse momentum and rapidity of outgoing back-to-back partons.
\Cref{fig:fullbjorkenx} shows the distribution of $x_{1,2}$ values for different event selections relevant for $I_{\rm AA}$. We used \texttt{Pythia8} with LO matrix element and EPPS21 NLO nPDFs (as LO ones are not available). We used the \texttt{Pythia8::Info::x1pdf()} function to access $x_{1,2}$ at the highest scale. We neglect the $x_{1,2}$ sampling during initial state radiation for simplification and better agreement with the strict LO picture. The left panel shows the probed $x_{1,2}$ values in the inclusive jet selection for $p_T^j>30\,\text{GeV}$. The steeply falling jet spectrum biases the $x$ values towards the bound controlled by $p_{T,\text{min}}$, \cref{eq:bjorkenx} (dotted line). The middle panel shows $h+j$ events for low momentum hadrons $p_T<10\,\text{GeV}$ that comes from fragmentation of $p_T\approx p_{T,\text{min}}$ jets. Therefore, the probed $x_{1,2}$ in these selections are similar to those probed by inclusive jets and nPDF uncertainties are positively correlated in \cref{fig:pearsonjet}. The rightmost panel is for high momentum selection (hadrons with $p_T>40\,\text{GeV}$). This time hadrons must come from a larger momentum jet with $p_T > p_{T,\text{min}}$. Correspondingly, a high momentum hadron cross section corresponds to larger values of Bjorken-$x$ and is less correlated with an inclusive jet cross section as seen in \cref{fig:pearsonjet}.

The difference in the probed Bjorken-$x$ by the trigger and semi-inclusive cross sections can be also used to explain the observed $I_{\rm OO}$ deviation from unity. Depending on the values of Bjorken-$x$, parton distributions inside the nuclei (and cross sections) are suppressed/enhanced compared with a free proton. This can be seen in the inclusive jet $R_{\rm OO}$, cf.,~\cref{fig:RAA_6800,fig:RAA_200}. For very small momentum (Bjorken-$x$), we expect the gluon shadowing to suppress the jet production, while for the intermediate momentum, it is enhanced by anti-shadowing. For large jet momentum (Bjorken-$x$), the parton distributions and, hence, cross sections are suppressed by the EMC effect. Such a rising and then falling trend is well seen for EPPS21 results in \cref{fig:RAA_6800}. For TUJU21 and nNNPDF3.0, we observe a mostly falling trend. Note that at lower collision energies shown in \cref{fig:RAA_200}, the probed Bjorken-$x$ is larger and we observe the decreasing  $R_{\rm OO}$ with momentum. The trigger jet cross section is dominated by jets of momentum just above the threshold $p_{T,\text{min}}^{j}$, therefore we can approximate $\left.\sigma^j\right|_{p_T^j> p_{T,\text{min}}^{j}} = \int_{p_{T,\text{min}}^{j}}^{\infty} \dd p_{T} \, \dd\sigma^j(p_T)/\dd p_T \approx \Delta p^j_{T,\text{min}} \dd\sigma^j(p^j_{T,\text{min}})/\dd p_T$. Using this, the semi-inclusive nuclear modification factor can be written as the ratio of two nuclear modification factors
\begin{equation}
I_\mathrm{AA}(p_T) \approx \frac{R^{h+j}_\mathrm{AA}(p_T)}{R_\mathrm{AA}^j(p^j_{T,\text{min}})} \;.
\end{equation}
As shown above,  $h+j$ cross section with hadron momentum above $p_{T,\text{min}}^{j}$ probes larger values of Bjorken-$x$ than the trigger. For EPPS21 nPDFs,  jet $R^{j}_\mathrm{OO}$ is increasing in the region $40\,\text{GeV}<p_T<200\,\text{GeV}$, while for TUJU21 it is constant or decreasing. We expect that hadrons with momentum $p_T<90\,\text{GeV}$ will be produced mostly by jets with  $p_T<200\,\text{GeV}$. Then $I_\mathrm{OO}$ will be larger or smaller than unity, depending if $R^{j}_\mathrm{OO}$ increases or decreases above the trigger momentum.
Indeed in \cref{fig:jet-hadronat6800} we see that for large hadron momentum $I_\mathrm{OO}>1$ for EPPS21 and $I_\mathrm{OO}<1$ for TUJU21 nPDFs, in agreement with our qualitative argument. nNNPDF3.0 jet nuclear modification factors remain centered around unity which results in relatively constant central $I_\mathrm{OO}^h$. $R^{j}_\mathrm{OO}$ has negative slope at $\sqrt{s_{\rm NN}}=200\,\text{GeV}$ shown in \cref{fig:RAA_200} for both nPDF sets and indeed we see strongly suppressed $I_\mathrm{OO}<1$ in \cref{fig:jet-hadronat200}.

\subsection{Scale, shower, and hadronization uncertainties} 
\label{ssec:pnpunc}
Besides the nPDF uncertainties, there exist additional sources of error that need to be considered. These include uncertainties coming from variations of factorization and renormalization scales $\mu_{R,F}$ as well as the chosen shower and hadronization model. 

\begin{figure*}
    \centering
    \includegraphics[width=0.49\textwidth]{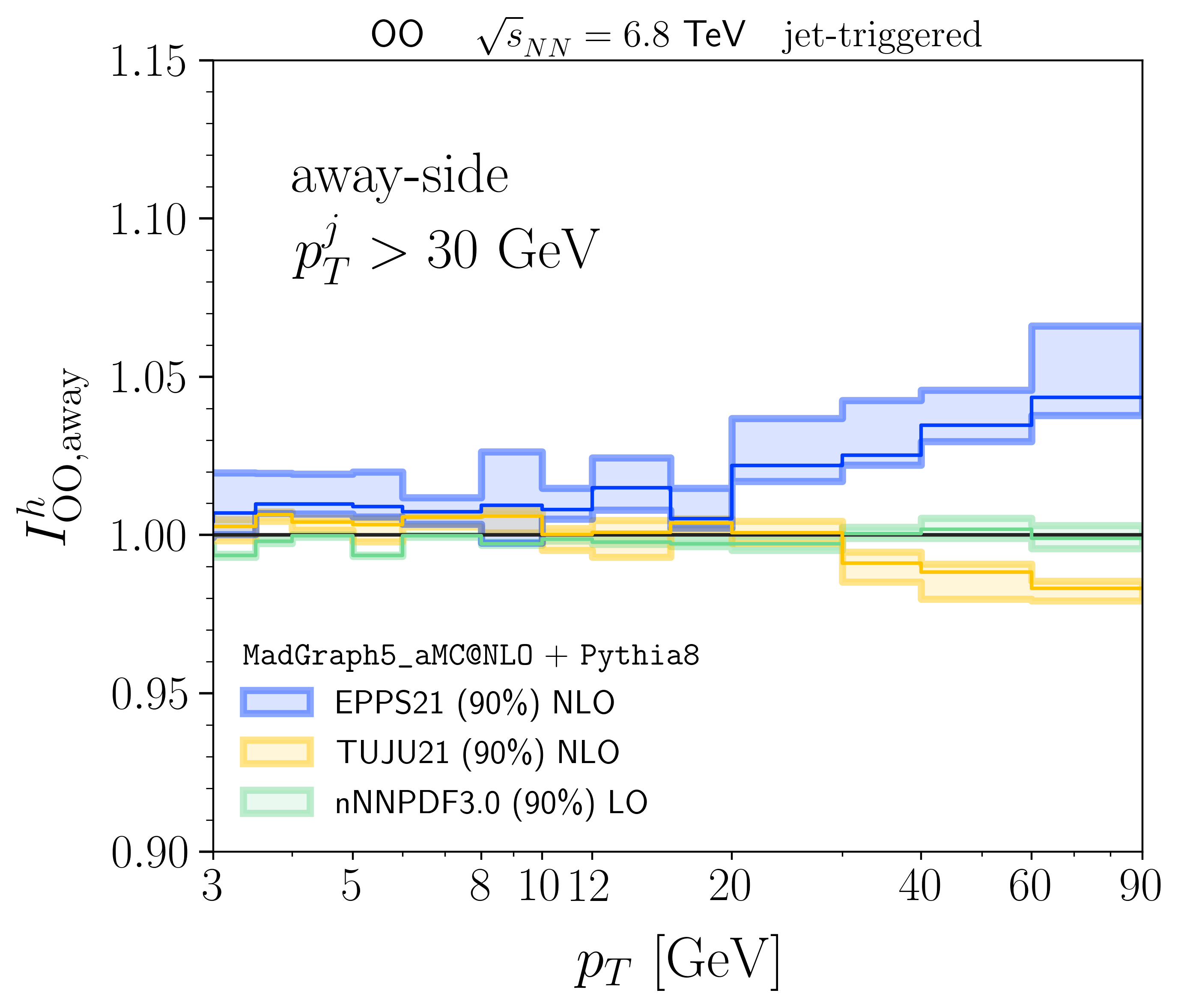}
    \caption{The NLO scale uncertainties in the no-quenching baseline for away-side jet-triggered hadron $I^{h}_\mathrm{AA}$  with $p_T^j > 30\,\text{GeV}$ in OO collisions at $\sqrt{s_\mathrm{NN}} = 6.8\,\text{TeV}$, cf.,~\cref{fig:jet-hadronat6800}.} 
    \label{fig:scale_unc}
\end{figure*}

The factorization and renormalization scales are not physical and the dependence of the cross section on them is the consequence of the truncated perturbative series. Therefore, varying $\mu_{R,F}$  is commonly used to estimate the uncertainties due to missing higher-order terms\footnote{Note that scale variations do not have a probabilistic interpretation of uncertainties. Instead, Bayesian models can be used to estimate the confidence intervals of missing higher-order terms~\cite{Duhr:2021mfd}.}.
We choose them to be equal to the sum of outgoing parton transverse momenta $\mu_{R,F} =  \frac{1}{2} \sum_i p_{T,i}$ and follow a 9-point variation technique to estimate the scale uncertainties by varying each of them by a factor of two up and down. In particular, we vary the scales correlatedly for all cross sections in $I_{\mathrm{AA}}$.
\Cref{fig:scale_unc} shows the scale uncertainties of the away-side $I_{\rm OO}$ for LHC kinematics\footnote{In this plot the central value of nNNPDF3.0 corresponds to the $0^\mathrm{th}$ member of the nPDF set which is the average over all replicas.}. They are around a few percent across all $p_T$, smaller than most nPDF uncertainty. In the low momentum region, nPDF and scale uncertainties are similar. Near-side hadrons, lower center-of-mass energies, and different $p_{T,\min}^j$ trigger cuts (not shown) have similar scale uncertainties.

\begin{figure*}
    \centering
    \includegraphics[width=0.50\textwidth]{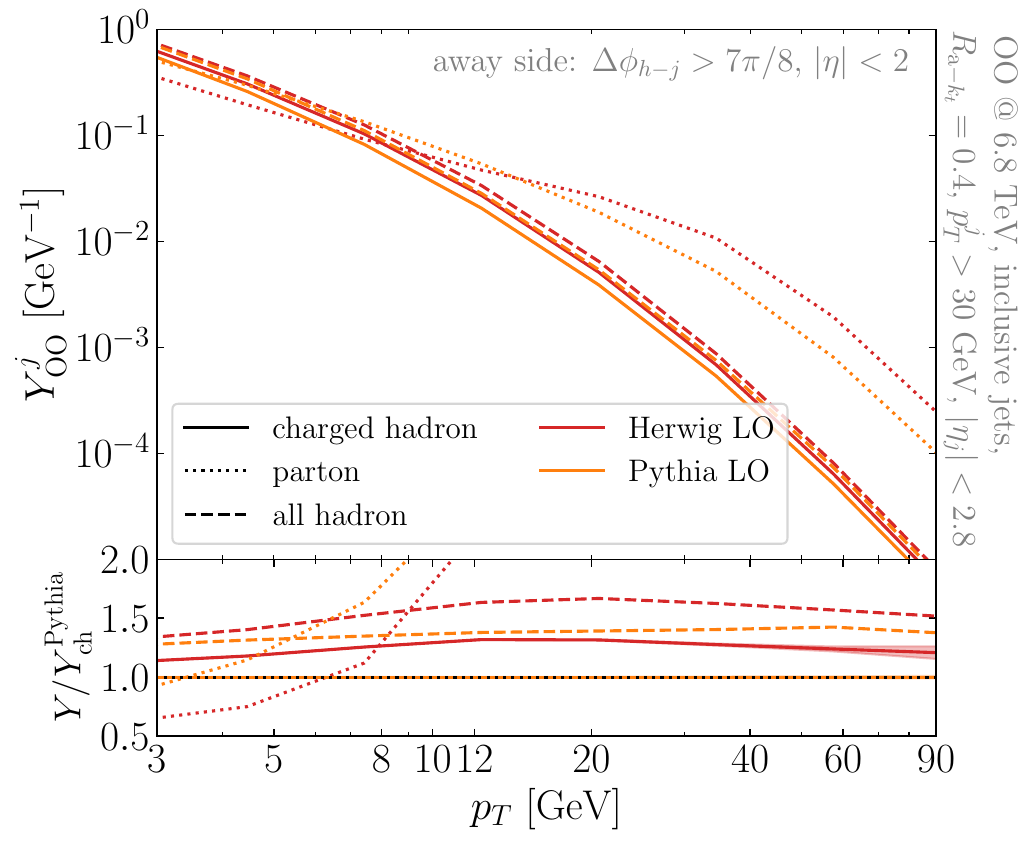}
    \includegraphics[width=0.48\textwidth]{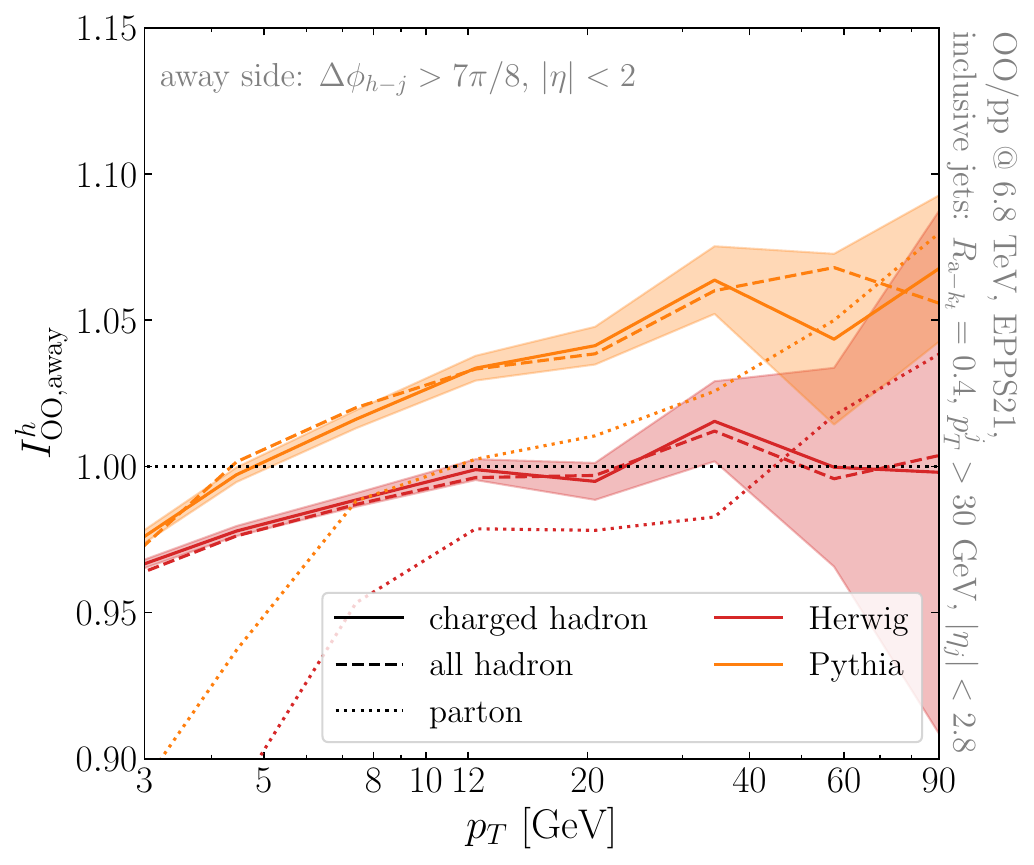}
    \caption{\textit{Left:} Jet triggered semi-inclusive particle spectrum  $Y_{\rm OO}$ normalised to Pythia charged. \textit{Right:} The jet triggered semi-inclusive nuclear modification factor $I_{\rm OO}$ for partons, charged and all hadrons. Bands indicate statistical uncertainties. 
    } 
    \label{fig:hadron_unc}
\end{figure*}

The trigger jet cross section is infrared and collinear safe (IRC), and robust with respect to the hadronization model. However, the distribution of charged hadrons can be sensitive to the details of the partonic shower and the non-perturbative hadronization effects. Since these are modeled phenomenologically, instead of a systematic uncertainty study, we estimate shower and hadronization dependence by comparing computations with two LO event generators: \texttt{Pythia8.312} and \texttt{Herwig7.3.0}, which use different shower and hadronization models. The left panel of \cref{fig:hadron_unc} shows $Y_{\rm OO}$ computed for partons, charged and all hadrons. Trigger jets were reconstructed on all hadrons or partons. The parton number is not an IRC-safe quantity. Partons fragment into lower momentum hadrons; therefore, partonic $Y_{\rm OO}$ is shifted to larger momentum compared with hadrons. The all-hadron $Y_{\rm OO}$ differs between Pythia and Herwig, roughly by a constant factor. Compared with the spectra including only charged hadrons, the all-hadron spectra are larger by a factor of $1.2$ - $1.3$. The trends are similar for away- and near-side  particles (not shown).

Finally, the right panel of \cref{fig:hadron_unc} shows $I_{\rm OO}$ with drastically reduced hadronization differences. We observed that there is no difference between charged and all-hadron  $I_{\rm OO}$. However, there is a systematic difference between Pythia and Herwig results, the latter being up to $5\%$ suppressed. In the low momentum region, the difference is similar in size to that of nPDF uncertainties. We note the positive slope in this region for $I_{\rm OO}$ in LO computations, which is absent in NLO computations, indicating that further studies of shower and hadronization uncertainties might be needed. Sensitivity to hadronization and shower models can be reduced by measuring recoiling jets instead of hadrons, i.e., dijets~\cite{ATLAS:2023zfx,ALICE:2023plt,ATLAS:2024jtu,Li:2024uzk}.

\section{Hadron-triggered jet nuclear modification factor $I_{\rm AA}^j$}\label{sec:IAAhadron}

\begin{figure} 
    \centering
    \includegraphics[width=0.49\textwidth]{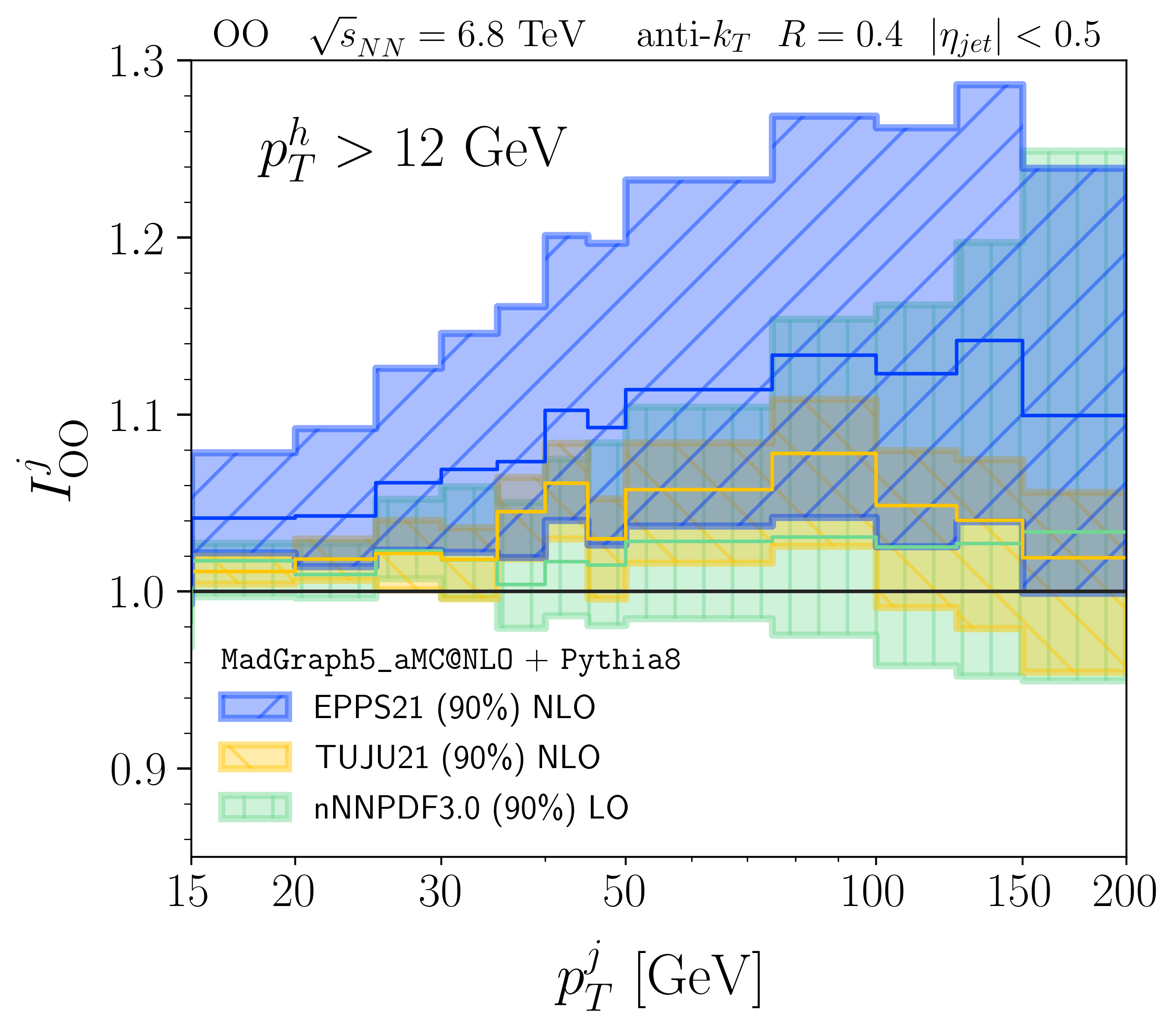}
    \includegraphics[width=0.49\textwidth]{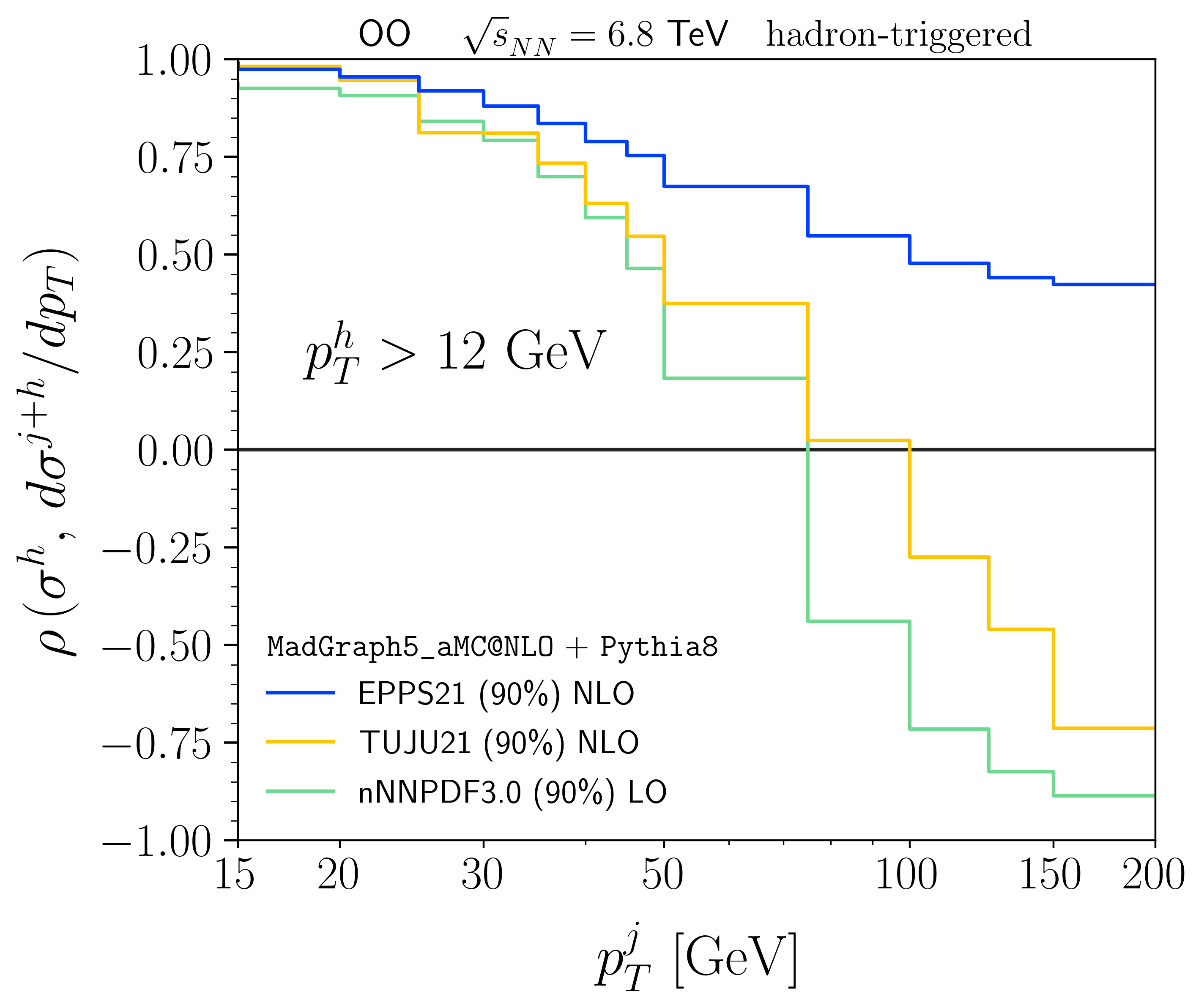}
    \caption{\textit{Left}: No-quenching baseline for hadron-triggered $I^{j}_\mathrm{AA}$ in oxygen-oxygen collisions at $\sqrt{s_\mathrm{NN}} = 6.8\,\text{TeV}$. The solid line corresponds to the central, whereas the bands show nPDF uncertainties at a $90\%$ confidence level. \textit{Right}: The Pearson correlation coefficient shows the nPDF error correlation of inclusive hadron cross section $\sigma^{h}$ and the hadron-jet cross section $\dd\sigma^{j+h}/\dd p_{T}$.} 
    \label{fig:hadron-jet12_6800}
\end{figure} 

An alternative strategy of making semi-inclusive measurements is counting jets recoiling from a high momentum charged hadron~\cite{ALICE:2015mdb,ALICE:2017svf,STAR:2017hhs,STAR:2023ksv}. Motivated by these studies, in this section, we study the nPDF uncertainties of hadron-triggered jet $I_{\rm AA}^j$. We require charged trigger hadrons to have a minimum transverse momentum $p_{T,\mathrm{min}}^{h}$ and we count charged jets that are opposite in azimuthal angle to the trigger hadron ($\Delta\Phi < \pi - 0.6$). The hadron with the highest transverse momentum is chosen as the trigger. The hadron-triggered jet yield reads 
\begin{equation}
    \label{eq:YAAh}
    Y_\mathrm{AA}^{j}(p_{T}^j) ~=~ \frac{1}{\sigma^{h}_\mathrm{AA}} \frac{\dd\sigma^{j+h}_\mathrm{AA}}{\dd p_{T}^j} ~ \Bigg\vert_{\,p_{T}^{h} > p_{T,\mathrm{min}}^{h}\, \;, ~ \Delta\Phi > (\pi - 0.6)} \, \;.
\end{equation}
Here, $\sigma^{h}_\mathrm{AA}$ is the inclusive hadron trigger cross section and $\dd\sigma^{j+h}_\mathrm{AA}/\dd p_{T}^j$ the corresponding cross section of finding a jet opposite to it. Jets are reconstructed from charged particles using the anti-$k_T$ algorithm with $R=0.4$. Trigger hadrons have $\vert \eta \vert < 0.9$ and charged jets $\vert \eta_\mathrm{jet} \vert < 0.5$. The semi-inclusive modification factor is obtained by the ratio
\begin{equation}
    I_\mathrm{AA}^j(p_T^j) = \frac{Y_\mathrm{AA}^{j}(p_T^j)}{Y_{pp}^{j}(p_T^j)} \;.\label{eq:Iaahad}
\end{equation}
We note here that experimental analyses~\cite{ALICE:2015mdb} use the difference of semi-inclusive jet spectra with trigger hadrons in two momentum ranges, i.e., $\Delta_\text{recoil}$ observable, to suppress the uncorrelated jet yield. Correspondingly, the ratio of these differences, $\Delta I_\mathrm{AA}^j(p_T^j)$, is used to measure medium-induced energy loss. For simplicity and computational efficiency, we use only a single lower momentum trigger $p_{T}^{h}>p_{T,\mathrm{min}}^{h}$ and show results for \cref{eq:Iaahad}.

No-quenching predictions of $I_\mathrm{OO}^j$ with $p_{T,\mathrm{min}}^{h} = 12\,\text{GeV}$ at $\sqrt{s_\mathrm{NN}} = 6.8\,\text{TeV}$ are presented in the left panel of \cref{fig:hadron-jet12_6800}. The EPPS21 and similarly nNNPDF3.0 nPDF uncertainties are growing with momentum from $5\%$ up to $15\%$.
The central values show significant deviations from unity ranging from $5\%$ to approximately $10\%$. Even within large nPDF uncertainties, the EPPS21 baseline shows deviations from unity. At lower $p_T^j$ TUJU21 nPDF uncertainties are around one percent and become roughly $5\%$ at larger momentum. Although central values of TUJU21 are closer to unity, deviations are again visible. 

A correlation analysis of the trigger cross section $\sigma^h$ and the coincidence cross section $\sigma^{j+h}$ in the form of the Pearson correlation coefficient is presented in the right panel of \cref{fig:hadron-jet12_6800}. While an almost perfect correlation is apparent in the lowest $p_T^j$ bin, it declines quickly for higher momenta. Cross sections obtained with TUJU21 and nNNPDF3.0 even become anti-correlated for $p_T^j > 75\,\text{GeV}$. The loss of correlation accompanies growing nPDF uncertainties for $I^j_\mathrm{OO}$. In comparison to jet-triggered results, correlations are lost more quickly.
This can be traced back to that the trigger hadron originates from a jet with slightly higher $p_T^j$, not far from the threshold, while jets on the other side with much higher momentum will probe significantly different Bjorken-$x$.
Therefore, the correlation is quickly lost for jets with momentum that is twice the threshold.

\begin{figure} 
    \centering
    \includegraphics[width=0.49\textwidth]{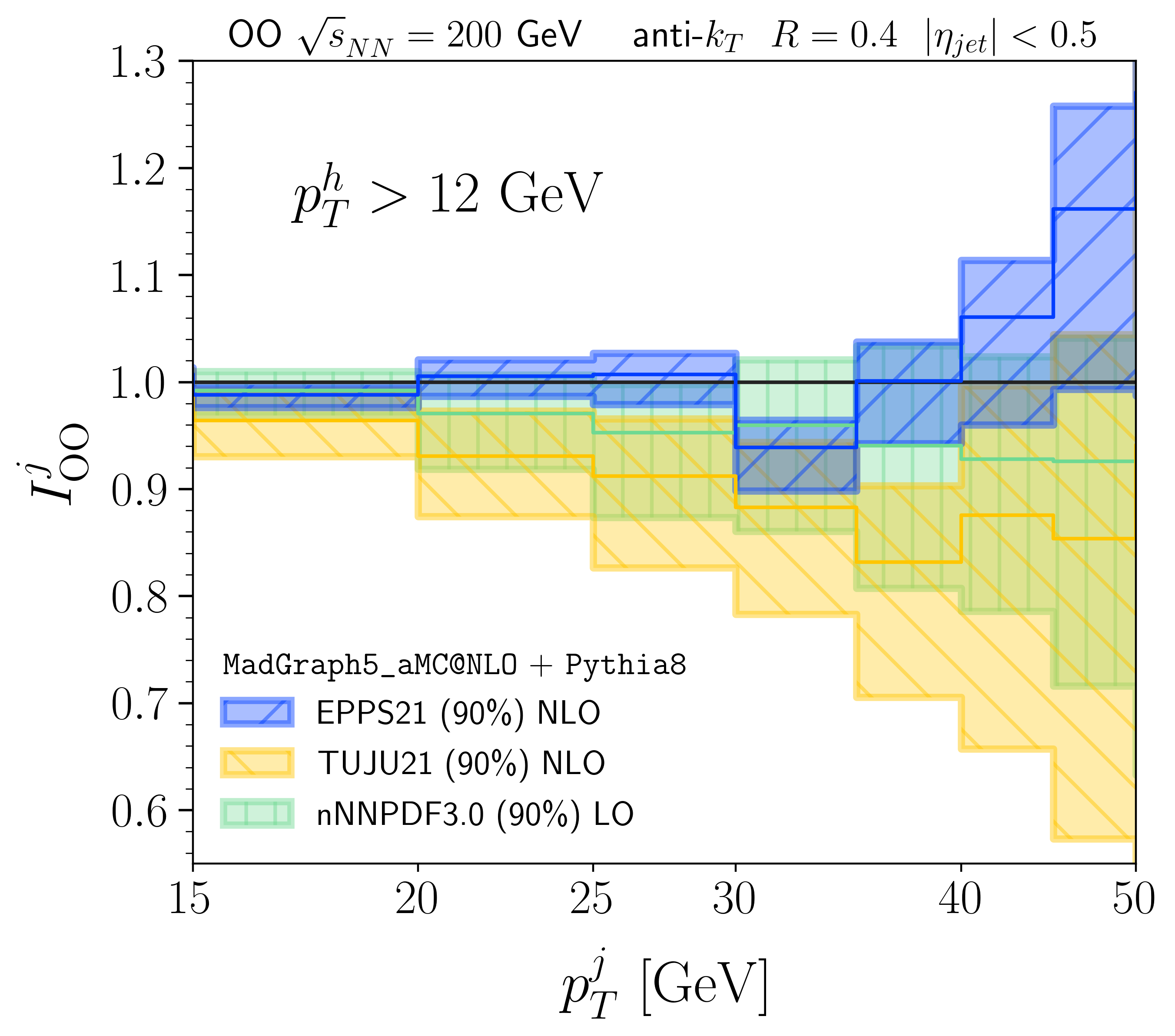}
    \caption{No-quenching baseline for hadron-triggered $I^{j}_\mathrm{AA}$ in oxygen-oxygen collisions at $\sqrt{s_\mathrm{NN}} = 200\,\text{GeV}$. The solid line corresponds to the central, whereas the bands show nPDF uncertainties at a $90\%$ confidence level.} 
    \label{fig:hadron-jet12_200}
\end{figure} 

Results at $\sqrt{s_\mathrm{NN}} = 200\,\mathrm{GeV}$, shown in \cref{fig:hadron-jet12_200}, agree with earlier observations. Again, we find that at lower collision energies, nPDF uncertainties using TUJU21  and nNNPDF3.0 are larger than for EPPS21. Overall, errors for hadron-triggered $I_\mathrm{OO}$ tend to be larger than for the jet-triggered $I_\mathrm{OO}$. We see that, in this case, there is a significant discrepancy between the baselines with EPPS21 and TUJU21 nPDFs, which emphasizes the need to consider multiple nPDF parametrizations for reliable baseline prediction.

\section{Conclusions and outlook}
\label{sec:conclusions}

The discovery of medium-induced suppression of high momentum jets, i.e., jet quenching, in small collision systems is an outstanding experimental challenge.
Large systematic uncertainties and biases have prevented such discovery in proton-nucleus and peripheral nucleus-nucleus collisions. Therefore, the collisions of light ions, namely, oxygen-oxygen, present a unique opportunity for such a discovery. In this work, we made the no-quenching baseline predictions for various energy loss observables in minimum bias OO collisions for (planned in 2025) LHC and (taken in 2021) RHIC measurements and exhaustively studied theoretical uncertainties. The statistically significant deviation between the experimental measurement and the baseline in OO collisions would be unambiguous proof of final state modification of high momentum particles in collisions with just around ten participating nucleons.

We studied in detail the inclusive jet nuclear modification factor $R_\mathrm{AA}^{j}$ and the jet-, and hadron-triggered semi-inclusive modification factors $I_{\rm AA}^h$ and $I_{\rm AA}^j$ in OO collisions. We evaluated the no-quenching baseline of these observables using state-of-the-art NLO matrix elements, NLO nPDFs, matched parton showers with hadronization and studied uncertainties arising from all of these components.

We find that the uncertainty in oxygen nPDFs leads to up to $10\%$ uncertainties on jet nuclear modification factor $R_\mathrm{OO}^{j}$ at LHC collision energies. The central value is consistent with unity within nPDF uncertainties in the studied momentum range. Scale, hadronization, and shower uncertainties are negligible compared with those from nPDFs. At RHIC collision energies, $R_\mathrm{OO}^{j}$ uncertainties are similar in size but with clear suppression at large jet momentum due to the nPDFs. Different nPDF extractions lead to different central values, but mostly consistent nPDF uncertainty bands. These uncertainties and deviations from  $R_{\rm OO}=1$ must be taken seriously when interpreting the magnitude of energy loss in measurements.

The nPDF uncertainties can be drastically reduced by considering a double ratio of cross sections, e.g., jet-triggered semi-inclusive $I_{\rm AA}$. However, we find such uncertainty cancellation only for recoiling hadrons below the jet trigger momentum. At higher momentum, the $I_{\rm OO}$ uncertainties increase substantially because the trigger jets and recoiling high momentum hadrons probe nPDFs at different Bjorken-$x$.  Although, at LHC, $I_{\rm OO}$ remains consistent with unity within uncertainty bands, central values of different nPDF sets can show opposing trends. This underlies the importance of systematically comparing several nPDFs for robust baseline computation. At RHIC energies, we observe a significant suppression (up to $20\%$) of $I_{\rm OO}$ for large hadron momentum due to nPDFs. We also show that scale uncertainties cancel significantly and predictions show little sensitivity to parton shower and hadronization models in the double ratio.

For hadron-triggered semi-inclusive jet $I_{\rm OO}$, the nPDF uncertainty is observed to cancel close to the hadron trigger momentum, but it grows rapidly at larger momenta. We note that at LHC energies hadron-triggered $I_{\rm OO}$ is above unity, even considering the nPDF uncertainties. Therefore unity clearly cannot be used as a baseline to quantify the potential energy loss signal (see also Ref.~\cite{He:2024rcv}). We note that at RHIC energies, predictions between different nPDF sets do not agree, which prevents us from making precise baseline predictions for these kinematics.

In conclusion, we presented state-of-the-art predictions for the no-quenching baseline for jet $R_\mathrm{AA}^{j}$ as well as jet-, and hadron-triggered $I_{\rm AA}$ observables with robust estimation of theoretical uncertainties along with model dependencies in oxygen-oxygen collisions\footnote{We expect near identical results in other light ion systems, e.g., neon-neon collisions.}. Whether an experimental discovery of energy loss in such a small system can be done will also crucially depend on the experimental precision of these measurements. With the expected integrated luminosity of $0.5\,\text{nb}^{-1}$~\cite{Brewer:2021kiv}, the statistics at LHC could be sufficient for a precise jet-triggered hadron $I^h_{\rm AA}$ measurement, see \cref{app:statunc}. Although minimum bias OO collisions do not have centrality selection uncertainty, $R_\mathrm{AA}$ and $I_\mathrm{AA}$-type measurements require a suitable $pp$ reference at the same collision energy. If a reference measurement at $\sqrt{s_\text{NN}}=6.8\,\text{TeV}$ is not available, alternative strategies of reference spectra interpolation or ratio of spectra at different energies could be considered~\cite{Brewer:2021tyv}. However, the estimation of experimental and theoretical uncertainties of such approaches is beyond the scope of the current paper.
Finally, a discovery is only possible if a sufficiently large signal is present. Although some predictions for $R_\mathrm{AA}$-type observables have been computed in the past~\cite{Huss:2020dwe,Zakharov:2021uza,Ke:2022gkq}, a systematic comparison of the sensitivity of different observables to medium modifications would be very valuable in guiding experimental and theoretical developments. We leave such computations for future work.

\begin{acknowledgments}
The authors thank 
Valentin Hirschi, 
Alexander Huss, 
Peter Jacobs, 
Petja Paakkinen,
and 
Juan Rojo,
for useful discussions.
This work is supported by the DFG through Emmy Noether Programme (project number 496831614) and CRC 1225 ISOQUANT (project number 27381115). We acknowledge support by the state of Baden-W\"urttemberg through bwHPC and the German Research Foundation (DFG) through grant no INST 39/963-1 FUGG (bwForCluster NEMO).
\end{acknowledgments}

\vspace{1.5em}

\noindent
\textbf{Code and Data Availability.} This article has associated code and data available on \url{https://zenodo.org/records/15052297}.

\appendix

\section{Event generator settings}\label{app:egsettings}
Cross sections are computed with the use of Monte Carlo event generators, more specifically \texttt{MadGraph5\_aMC@NLO}-\texttt{v.2.9.18}~\cite{Alwall:2014hca} for the generation of partonic cross sections at the level of the matrix elements that are matched to \texttt{Pythia8.306}~\cite{Sjostrand:2014zea,Bierlich:2022pfr} which performs the final state evolution, including parton showers and hadronization. 

In \texttt{MadGraph5\_aMC@NLO}, we use \texttt{model loop\_sm-no\_b\_mass} which sets the number of massless quark flavors to 5. Strong partonic scattering processes at NLO are generated with the command \texttt{generate p p > j j [QCD]} where \texttt{p} refers to the colliding proton/oxygen, and \texttt{j} refers to the outgoing (massless) parton species. The matrix element at NLO includes tree level dijet diagrams (LO), one-loop corrections to the tree level, and real corrections, e.g., tri-jet production. All outgoing partons are clustered into jets using the anti-$k_T$ algorithm with $R = 0.7$ (final state particles are re-clustered using a smaller jet-radius). In order to restrict the volume of phase space that is integrated over, we apply a lower cut on the transverse momentum of outgoing partonic jets during the generation of matrix elements by choosing the value of \texttt{ptj} in the MG5 \texttt{run\_card.dat}. Generally, this cut needs to be well below the kinematics that are relevant for the computed cross sections. Following Ref.~\cite{Frixione:1997ks,Frederix:2016ost}, at NLO it is necessary to employ a different momentum cut on the hardest (largest $p_T$) outgoing parton jet and all other outgoing ones. In particular, we used the following momentum cuts:

\begin{center}
\begin{tabular}{ |c|c|c|c|  }
\hline
 $\sqrt{s_\mathrm{NN}} ~ [\mathrm{TeV}]$ & Observable & $p_{T,\mathrm{cut}}^\mathrm{hard} ~ [\mathrm{GeV}]$ & $p_{T,\mathrm{cut}}^\mathrm{other} ~ [\mathrm{GeV}]$\\ [0.2em]
 \hline\hline
    
    \multirow{3}{*}{$0.2$} & $R^j_\mathrm{AA}$ & $8$ & $5$ \\
    & $I^h_\mathrm{AA}$ & $8$ & $5$ \\
    & $I^j_\mathrm{AA}$ & $8$ & $5$ \\ \hline

    \multirow{2}{*}{$5.02$} & $\sigma^j_\mathrm{AA}$ & $17$ & $14$ \\
    & $Y^h_\mathrm{AA}$ & $17$ & $14$ \\
 \hline
    \multirow{3}{*}{$6.8$} & $R^j_\mathrm{AA}$ & $17$ & $14$ \\
    & $I^h_\mathrm{AA}$ & $17$ & $14$ \\
    & $I^j_\mathrm{AA}$ & $10$ & $7$ \\ \hline
\end{tabular}
\end{center}
Additionally, an upper transverse momentum cut of $2000\,\mathrm{GeV}$ was chosen. A more efficient phase space sampling was achieved by reweighting events according to a user-specific biasing function. To balance the steeply falling jet spectrum, the sampling distribution is rescaled $\propto p_{T,\mathrm{max}}^{\alpha_\mathrm{bias}}$. Here, $p_{T,\mathrm{max}}$ is the transverse momentum of the hardest partonic jet and $\alpha_\mathrm{bias}$ the biasing power. It is typically chosen around 4, approximating the $p_T$-dependence of the jet spectrum. 

Interfacing and matching the generated partonic events to \texttt{Pythia8} is done automatically. Here, default MG5 settings were used except for slightly customized shower parameters. To perform a more realistic shower simulation, we used the following changes:

\begin{itemize}[leftmargin=60pt,itemsep=-5pt]
    \item[] \makebox[6cm]{\texttt{BeamRemnants:primordialKT}\hfill} \texttt{= on}
    \item[] \makebox[6cm]{\texttt{TimeShower:QEDshowerByGamma}\hfill} \texttt{= on}
    \item[] \makebox[6cm]{\texttt{TimeShower:QEDshowerByQ}\hfill} \texttt{= on}
    \item[] \makebox[6cm]{\texttt{TimeShower:QEDshowerByL}\hfill} \texttt{= on}
    \item[] \makebox[6cm]{\texttt{TimeShower:alphaEMorder}\hfill} \texttt{= 1}
    \item[] \makebox[6cm]{\texttt{SpaceShower:QEDshowerByQ}\hfill} \texttt{= on}
    \item[] \makebox[6cm]{\texttt{SpaceShower:QEDshowerByL}\hfill} \texttt{= on}
    \item[] \makebox[6cm]{\texttt{SpaceShower:alphaEMorder}\hfill} \texttt{= 1}
\end{itemize}

\section{Comparing computations at NLO and LO}\label{app:NLOvsLO}

\begin{figure}[h] 
    \centering
    \includegraphics[width=0.49\textwidth]{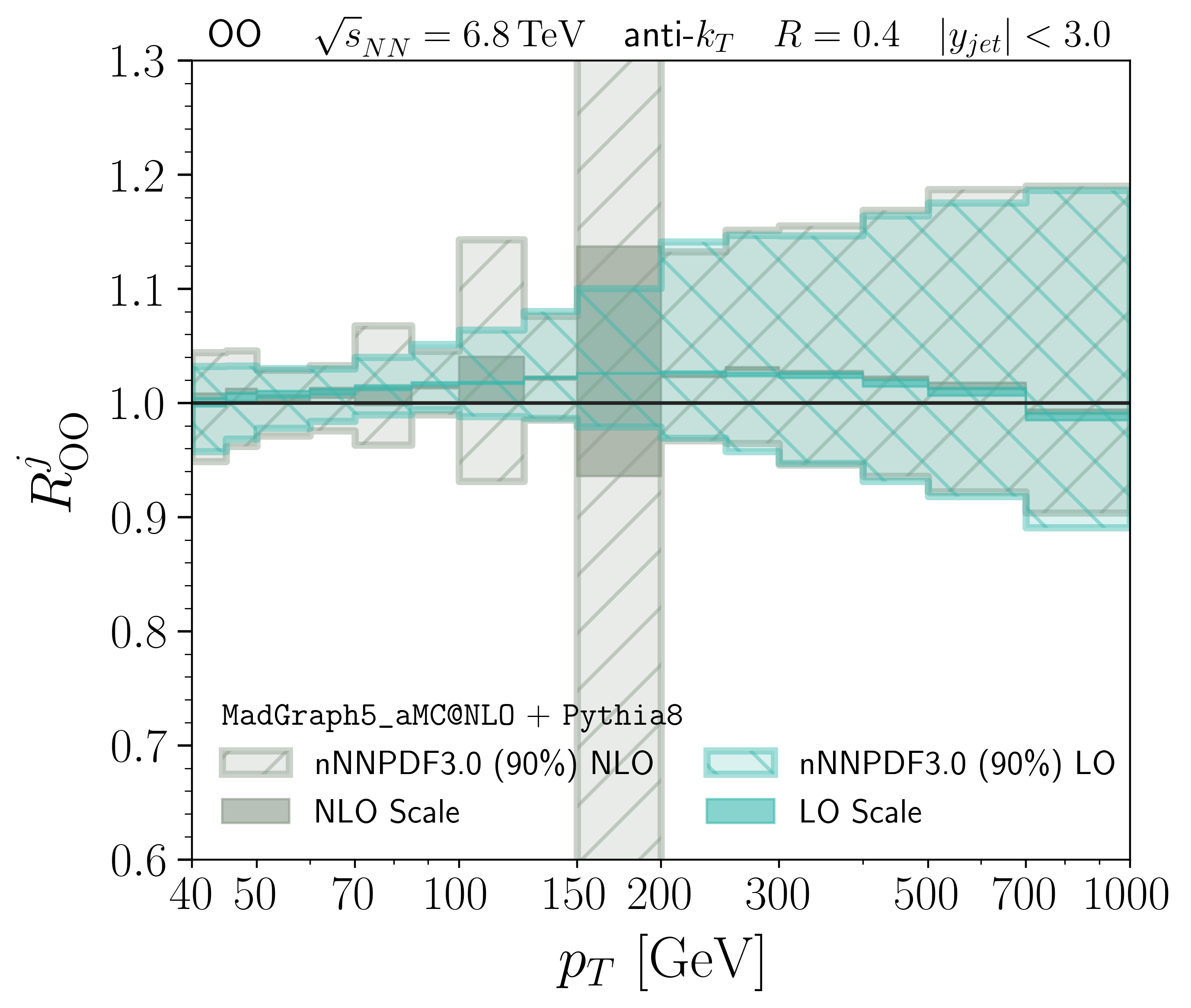}
    \includegraphics[width=0.49\textwidth]{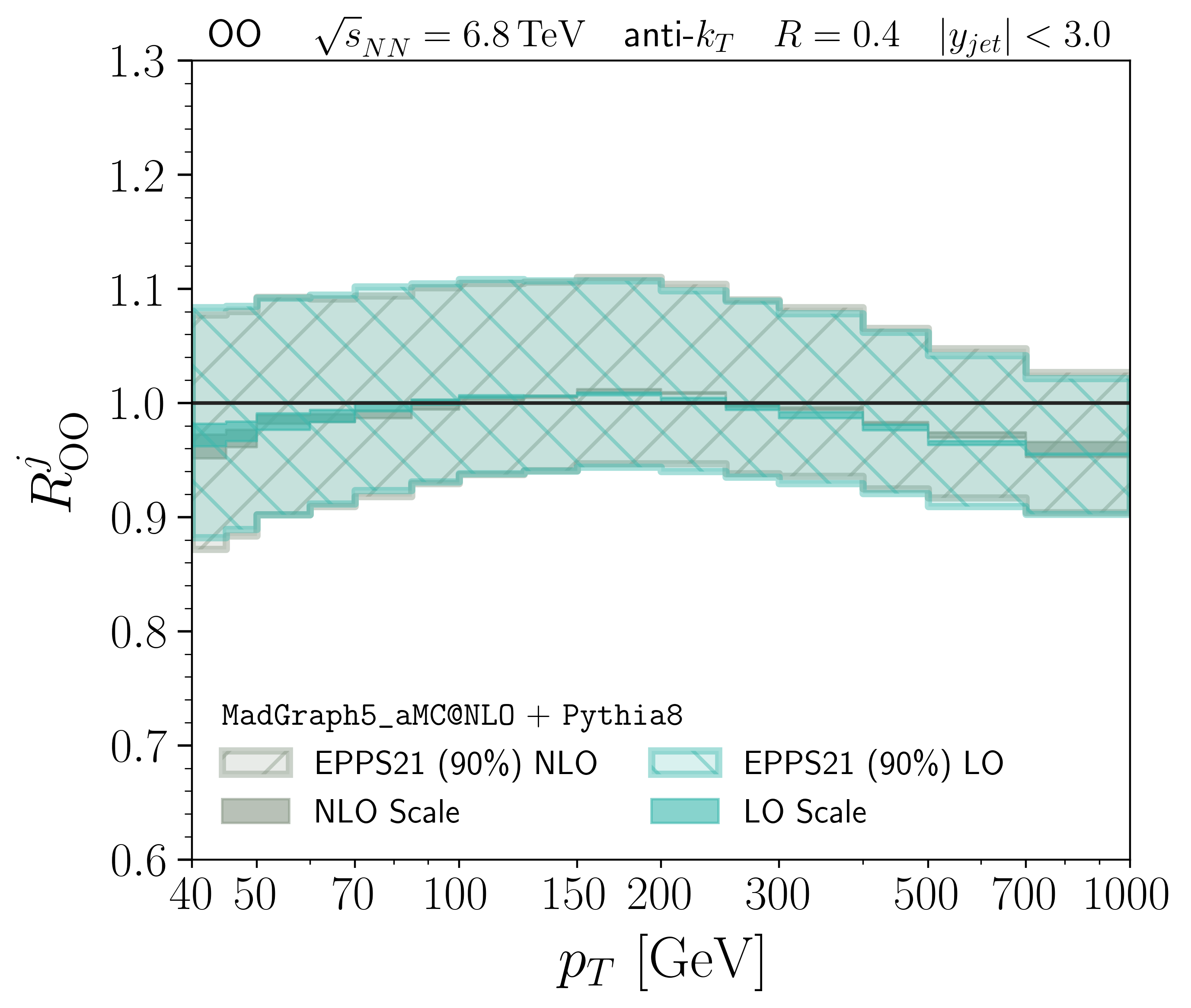}
    \caption{Comparing NLO with LO calculations of inclusive jet $R^j_\mathrm{OO}$ using nNNPDF3.0 (\textit{left}) and EPPS21 (\textit{right}) nPDFs at $\sqrt{s_\mathrm{NN}} = 6.8\,\mathrm{TeV}$.} 
    \label{fig:compNLOvsLO_RAA}
\end{figure}

\begin{figure}[h] 
    \centering
    \includegraphics[width=0.49\textwidth]{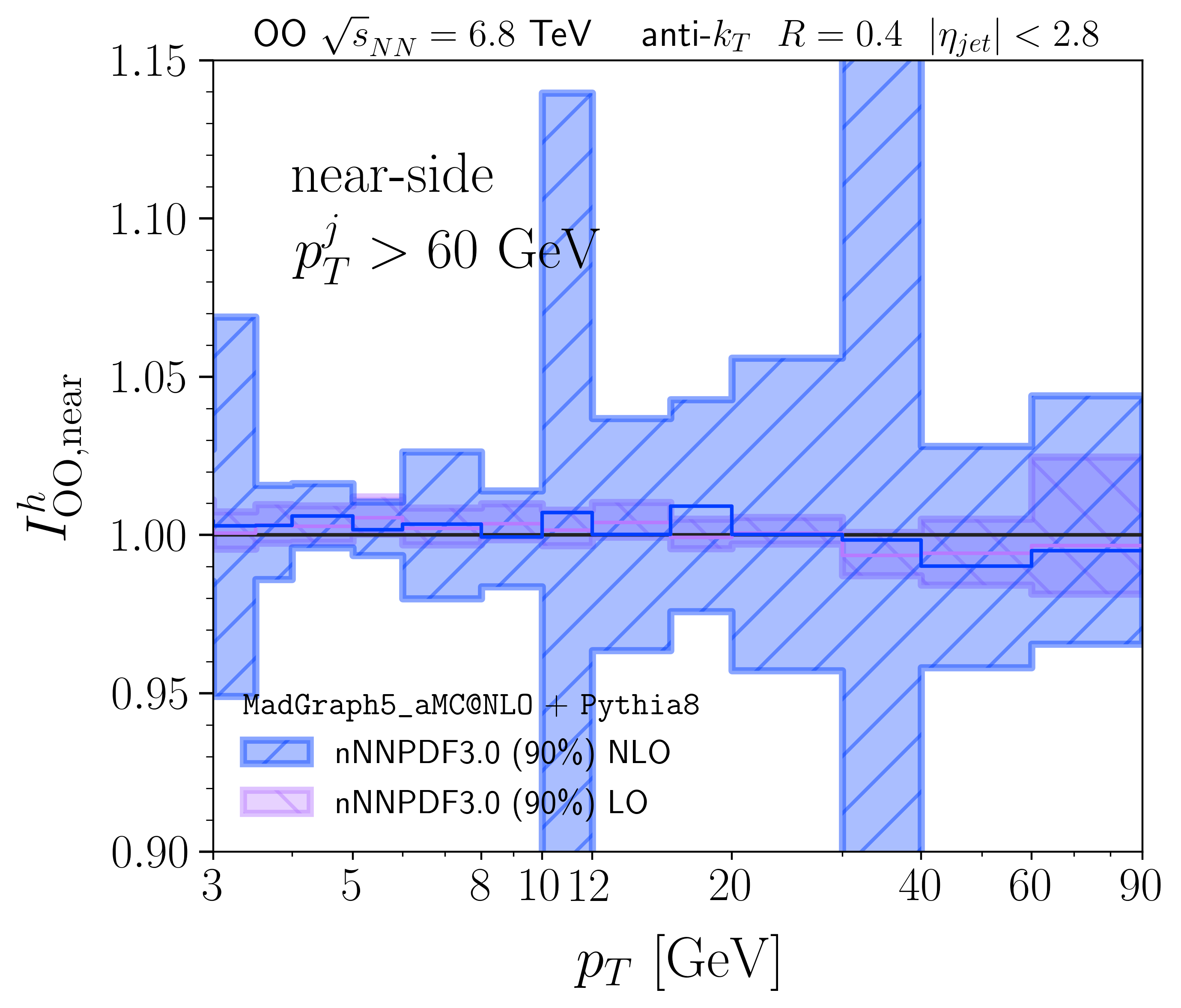}
    \includegraphics[width=0.49\textwidth]{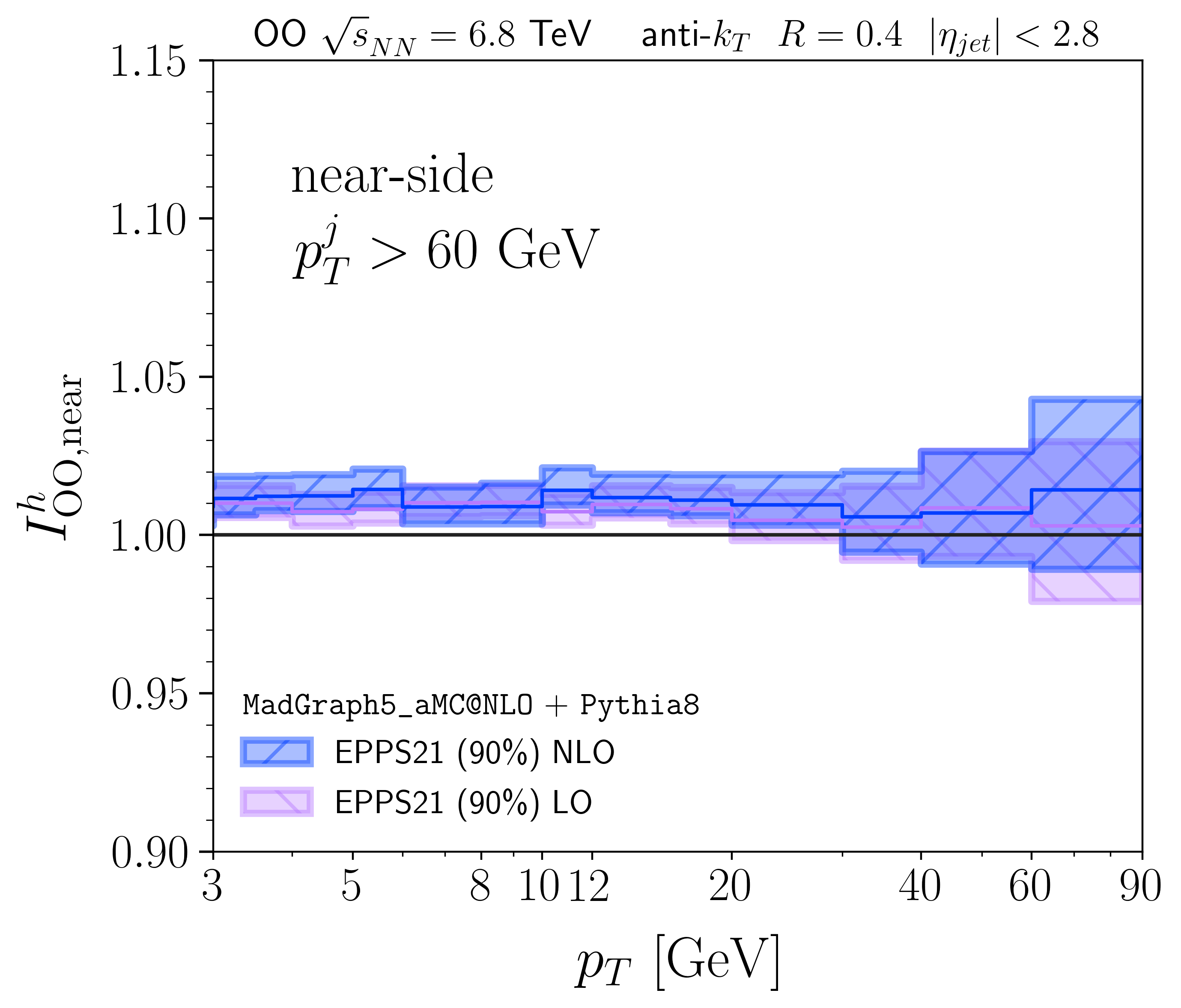}   
    \caption{Comparing NLO with LO calculations of semi-inclusive jet-triggered $I^h_\mathrm{OO}$ using nNNPDF3.0 (\textit{left}) and EPPS21 (\textit{right}) nPDFs at $\sqrt{s_\mathrm{NN}} = 6.8\,\mathrm{TeV}$.} 
    \label{fig:compNLOvsLO_IAA}
\end{figure}

In some cases uncertainty computations at NLO with \texttt{MadGraph\_aMC@NLO} can exhibit irregular behavior, leading to significant uncertainty increase. This phenomenon is illustrated in the left panel of \cref{fig:compNLOvsLO_RAA}, where we compare results for $R^j_\mathrm{OO}$ calculated using the NLO nNNPDF3.0 with both NLO and LO matrix elements. In particular, in certain $p_T$ bins both nPDF and scale uncertainties are significantly larger when using NLO matrix elements.
Generally, the effect of going to NLO on central values and nPDF uncertainties are minor for the inclusive jet nuclear modification factor in \cref{fig:compNLOvsLO_RAA} as well as semi-inclusive observables in \cref{fig:compNLOvsLO_IAA} where we representatively show results for $I^h_\mathrm{OO}$.
Since we did not manage to fully resolve the issue of irregular uncertainty behavior, and we consider it having numerical origin, when using nNNPDF3.0 in the main text, we show results with LO matrix elements.

\section{Absolute inclusive and semi-inclusive spectra}\label{app:absspectra}

This appendix presents comparisons of numerical predictions with experimental measurements in $pp$ collisions. Here, we use the same numerical setup as for the results shown in the main body of this article. This setup consists of NLO matrix element calculations using \texttt{MadGraph5\_aMC@NLO} matched to \texttt{Pythia8} to obtain showered and hadronized particle spectra (see \cref{app:egsettings} for details). 

\begin{figure*} 
    \centering
    \includegraphics[width=0.49\textwidth]{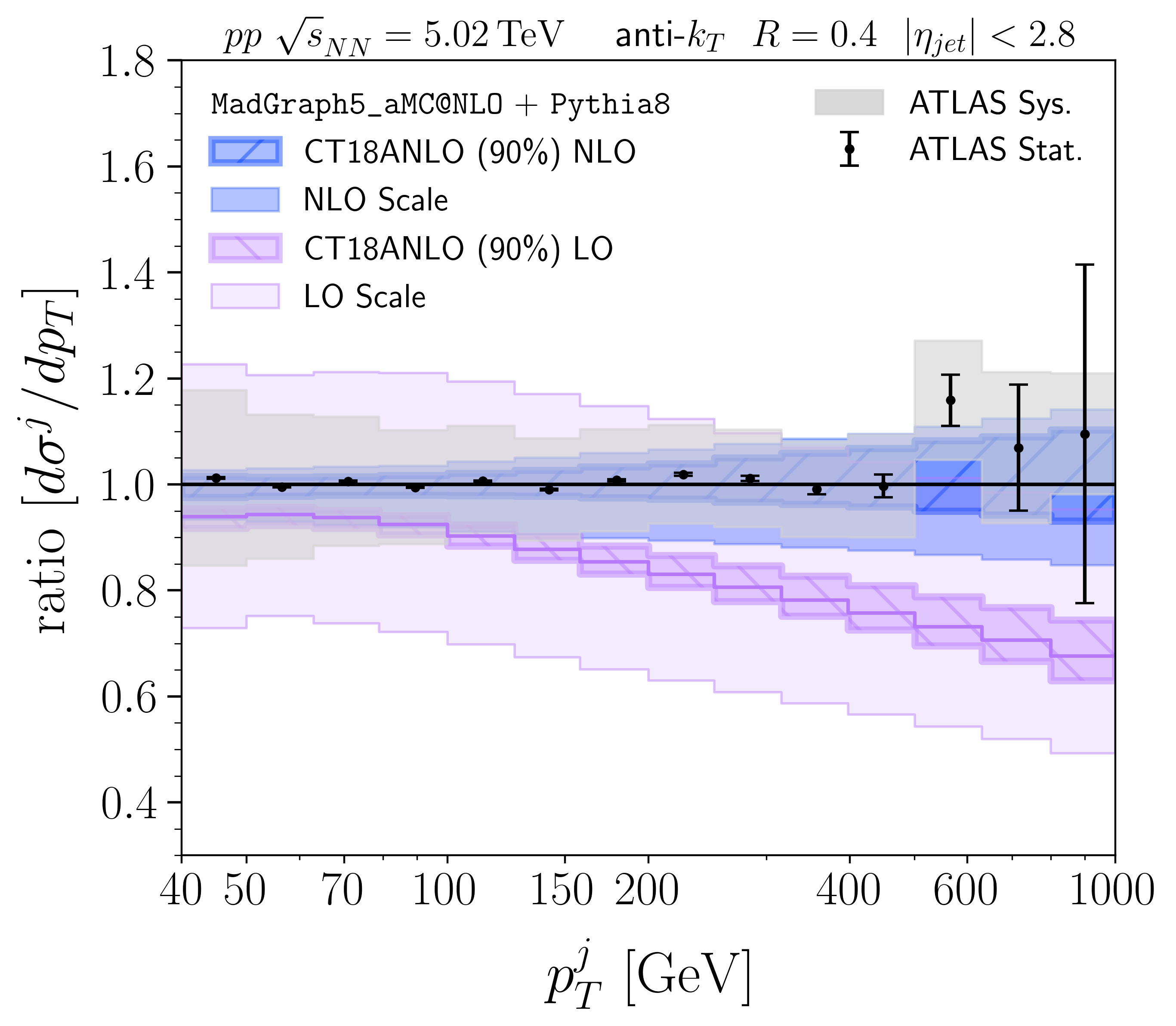}
    \caption{Comparing NLO and LO calculations of inclusive jet production at $\sqrt{s_\mathrm{NN}} = 5.02\,\text{TeV}$ done with \texttt{MadGraph5\_aMC@NLO} with $N_{f} = 5$ massless quark flavors in combination with \texttt{PYTHIA8} to experimental data from ATLAS \cite{ATLAS:2018gwx}. Jets were selected using the anti-$k_{T}$ and $R = 0.4$. Solid bands represent nPDF uncertainties (hatched) and scale uncertainties (unhatched). Scale uncertainty is estimated by scale variations of $\mu_{r}$ and $\mu_{f}$ with factors of $0.5$ and $2$. Error bars and bands around experimental data points correspond to statistical and systematic uncertainties.} 
    \label{fig:compINCATLAS}
\end{figure*}

In \cref{fig:compINCATLAS}, inclusive jet spectra $\dd\sigma/\dd p_T$ in $pp$ collisions at $\sqrt{s_\mathrm{NN}} = 5.02\,\mathrm{GeV}$ computed at NLO as well as LO are compared to experimental data from ATLAS~\cite{ATLAS:2018gwx}. We find excellent agreement between the experiment and NLO results. LO spectra become smaller than NLO ones as the jet momentum grows. As expected, going to NLO significantly decreases scale uncertainties. 

\begin{figure}[H] 
    \centering
    \includegraphics[width=0.49\textwidth]{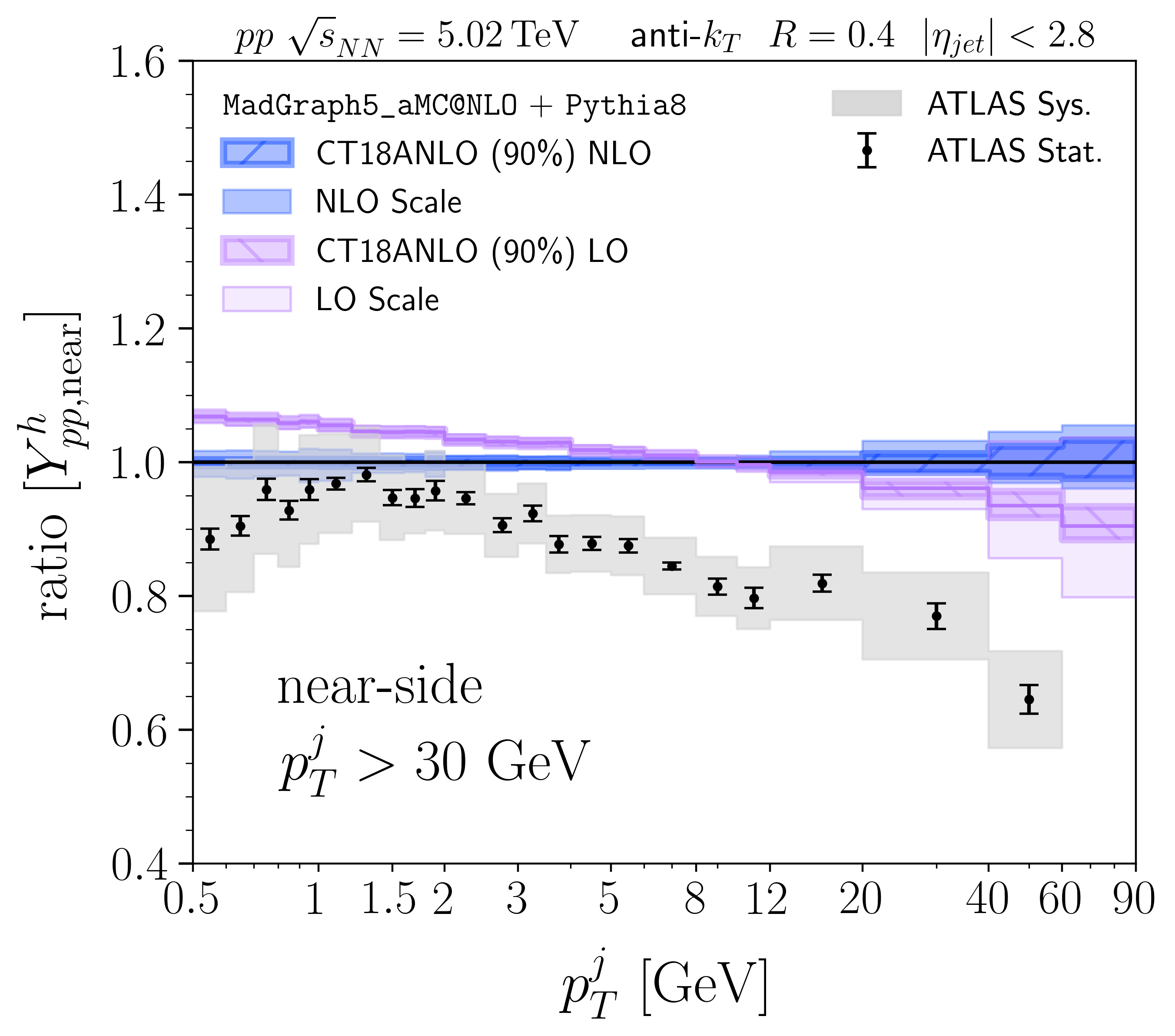}
    \includegraphics[width=0.49\textwidth]{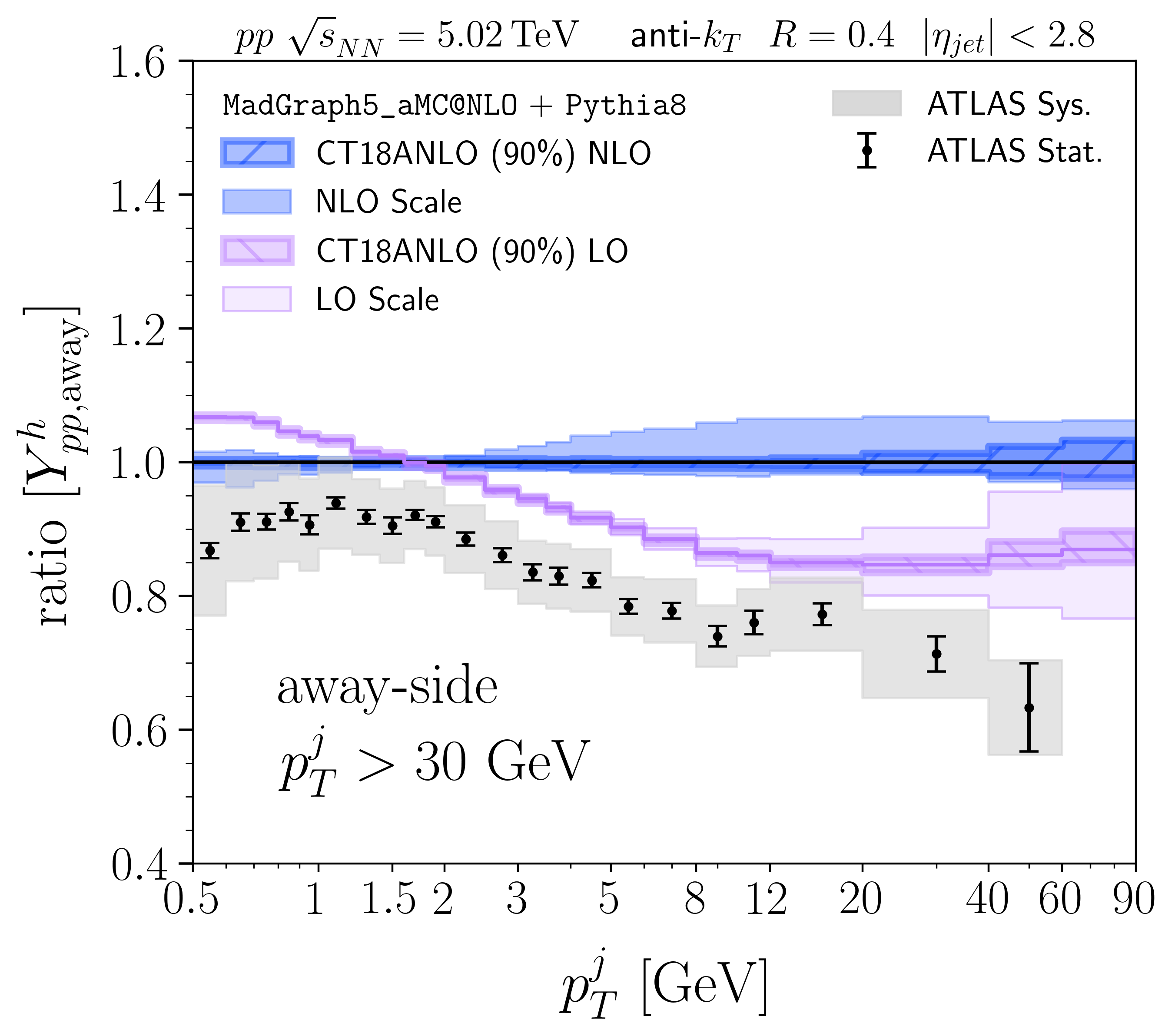}
    \caption{Comparing the jet-triggered hadron yield $Y^{h}_{pp}(p_{T})$ computed with \texttt{MadGraph5\_aMC@NLO} and \texttt{PYTHIA8} at NLO (purple) to the corresponding LO (blue) calculation and experimental data from ATLAS \cite{ATLAS:2022iyq}. Yields were divided by the NLO spectrum. The proton-proton collisions at $\sqrt{s_\mathrm{NN}} = 5.02\,\text{TeV}$ were simulated with the use of the CT18ANLO PDF set. The ratio of near-side yields for jets with $p_{T}^{j} > 30\,\text{GeV}$ is shown on the left and the right plot shows the corresponding away-side ratio. Solid bands represent nPDF uncertainties (hatched) and scale uncertainties (unhatched). Scale uncertainty is estimated by scale variations of $\mu_{r}$ and $\mu_{f}$ with factors of $0.5$ and $2$. Error bars and bands around experimental data points correspond to statistical and systematic uncertainties.} 
    \label{fig:compNLOLOATLAS}
\end{figure}

After demonstrating that the setup is well capable of reproducing inclusive observables, \cref{fig:compNLOLOATLAS} shows its performance for jet-triggered $Y_\mathrm{OO}^h$ using $p_{T,\mathrm{min}} = 30\,\mathrm{GeV}$, defined in \cref{eq:YAAjetaway} and \cref{eq:YAAjetnear}. For $p_T < 2\,\mathrm{GeV}$, the numerical predictions agree well with the experiment. Above $2\,\mathrm{GeV}$, the computation overpredicts the number of produced hadrons. The overproduction is less severe at LO than at NLO. Consequently, the LO predictions are closer to experimental observation while still not in agreement. When computing the ratio $Y_\mathrm{AA}/Y_{pp} = I_\mathrm{AA}$, the difference in magnitude cancels. Near-side and away-side comparisons (left and right columns) show identical characteristics.

\FloatBarrier

\section{Projection of statistical uncertainties}\label{app:statunc}

In this section, we estimate statistical uncertainties of jet-, and hadron-triggered semi-inclusive modification factors $I_{\rm AA}^h$ and $I_{\rm AA}^j$ in OO collisions. We assume perfect detector efficiency and the integrated OO luminosity of $\int \dd t \, L_{\rm AA} = 0.5\,\text{nb}^{-1}$ at $\sqrt{s_\mathrm{NN}}=6.8\,\text{TeV}$~\cite{Brewer:2021kiv}
and $\int \dd t \, L_{\rm AA} = 0.32\,\text{nb}^{-1}$ (400 million minimum bias events) at $\sqrt{s_\mathrm{NN}}=200\,\text{GeV}$~\cite{Liu:2022jtl}.

The relative uncertainty of a cross section $\sigma_\mathrm{AA}$ is estimated by 
\begin{equation}
    \frac{\Delta \sigma_\mathrm{AA}}{\sigma_\mathrm{AA}} = \frac{1}{\sqrt{N_\mathrm{AA}}} = \frac{1}{\sqrt{\mathcal{L}_\mathrm{AA} \sigma_\mathrm{AA}}} \,,
\end{equation}
where it assumed that the events that contribute follow Poisson statistics. $N_\mathrm{AA}$ is the total number of events and $\mathcal{L}_\mathrm{AA}$ the time-integrated luminosity. 

\begin{figure}[h] 
    \centering
    \includegraphics[width=0.49\textwidth]{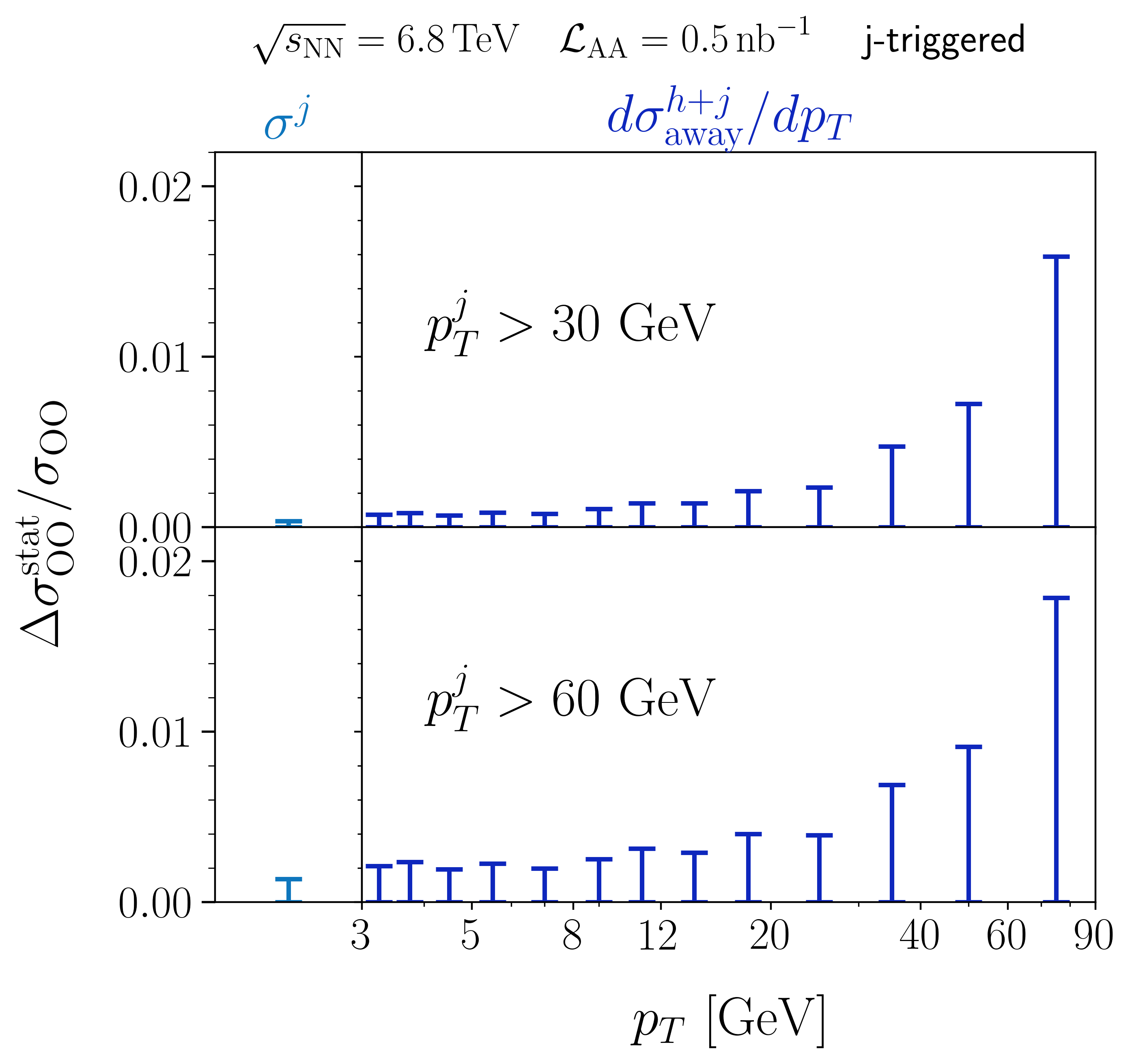}
    \includegraphics[width=0.49\textwidth]{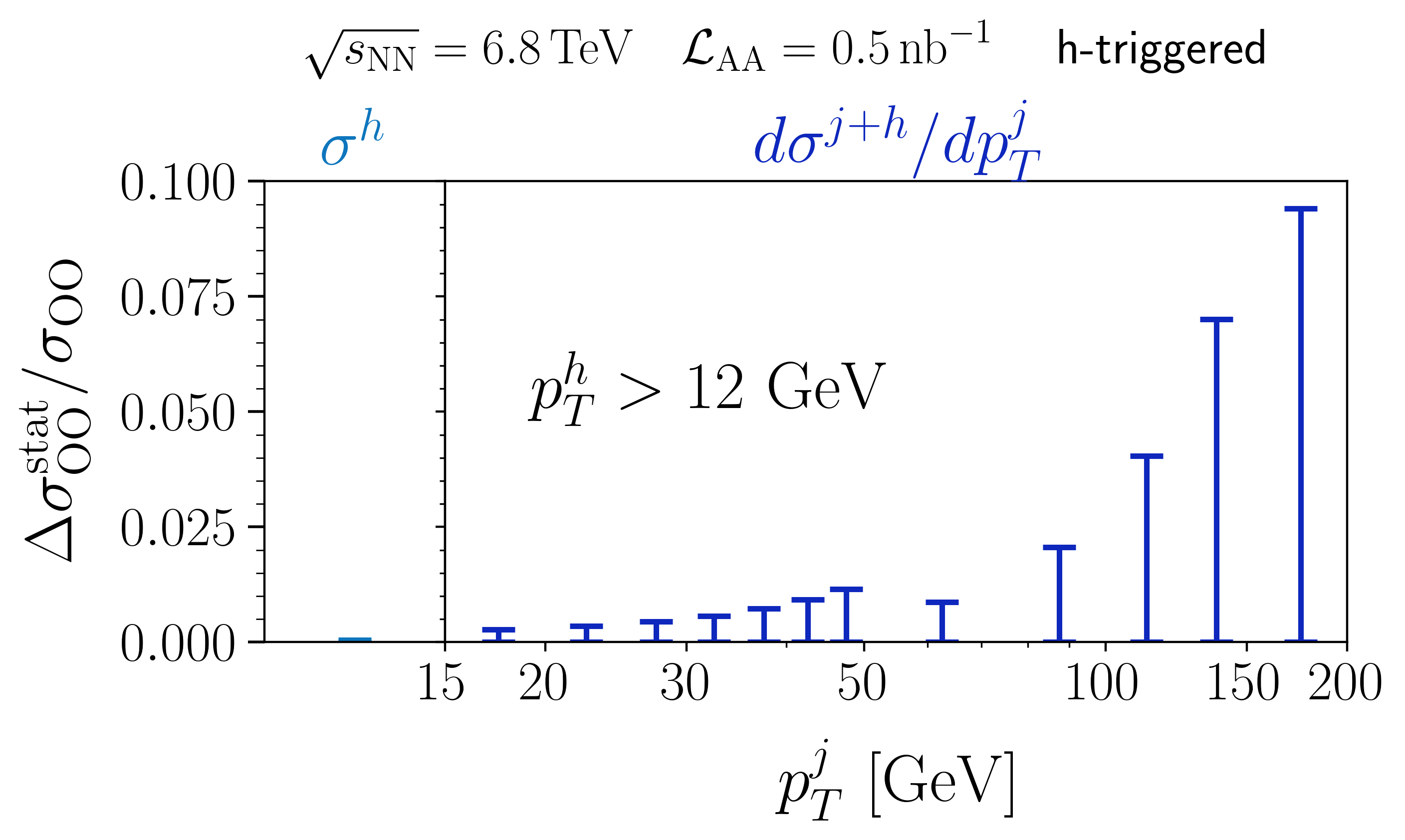}
    \caption{Projection of relative statistical uncertainties for trigger and coincidence cross sections at $\sqrt{s_\mathrm{NN}} = 6.8\,\mathrm{TeV}$ using cross sections obtained with EPPS21. \textit{Left}: Jet-triggered measurement with trigger momentum cutoff $p_{T,\mathrm{min}}^j=30\,\mathrm{GeV}$ (upper panel) and $60\,\mathrm{GeV}$ (lower panel). Results are shown for trigger cross section $\sigma^j$ and away-side $\dd\sigma^{h+j}/\dd p_T$. \textit{Right}: Hadron-triggered measurement with $p_{T,\mathrm{min}}^h=12\,\mathrm{GeV}$. Results are shown for trigger cross section $\sigma^h$ and $\dd\sigma^{j+h}/\dd p_T^j$.} 
    \label{fig:statuncatlas}
\end{figure}

\cref{fig:statuncatlas} shows the projected relative statistical uncertainty of jet-triggered and hadron-triggered cross sections for oxygen-oxygen collisions at LHC energies. The integrated beam luminosity is obtained from the ``moderately optimistic" running scenario in Ref.~\cite{Citron:2018lsq} and corresponds to the target luminosity \cite{Brewer:2021kiv}. Cross sections for jet-triggered measurements are expected to have well-bound statistical uncertainties that reach the percent level at the highest momentum bins. In comparison, the relative error of the coincidence cross section for hadron-triggered measurements reaches roughly $10\%$. 

\begin{figure}[h] 
    \centering
    \includegraphics[width=0.49\textwidth]{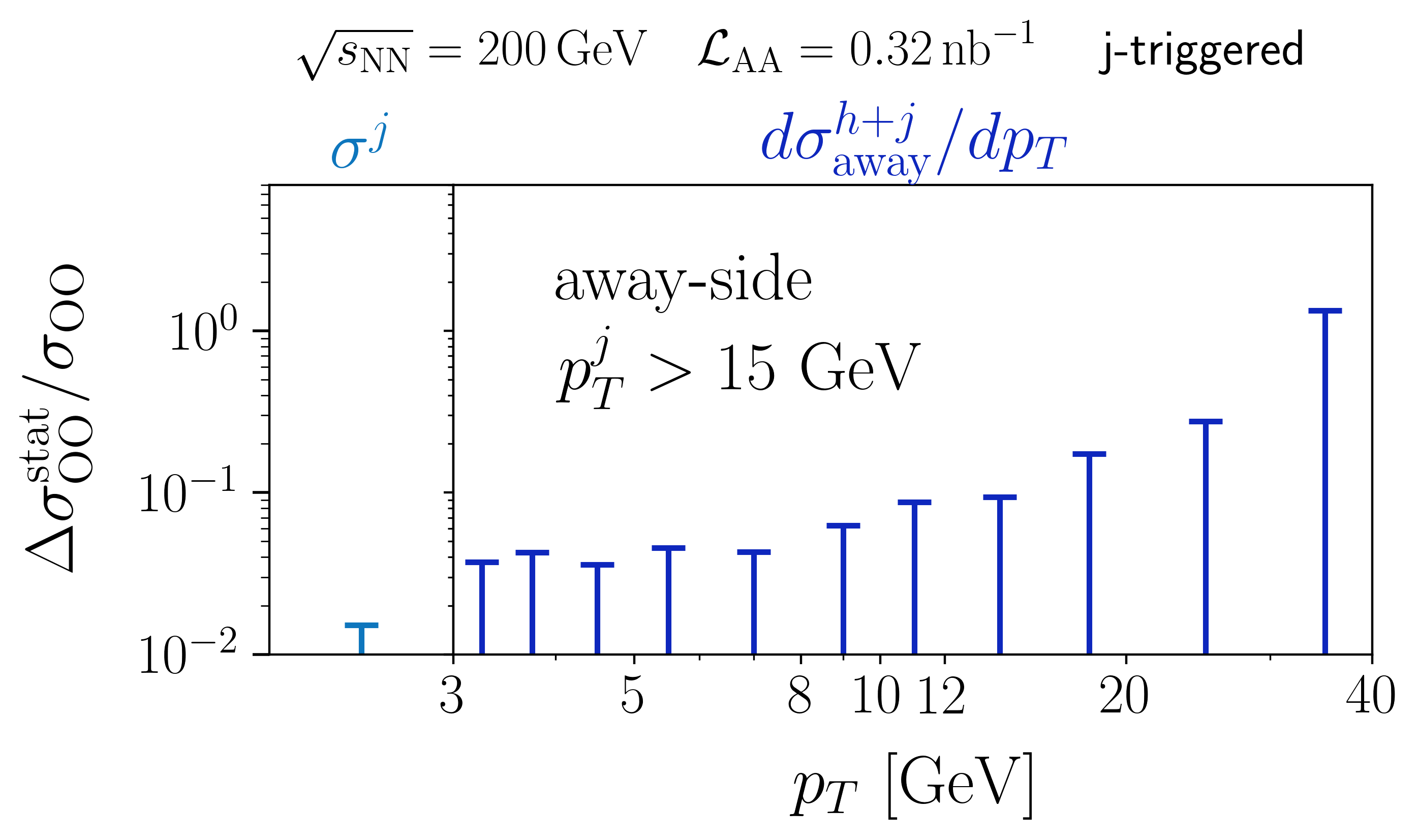}
    \includegraphics[width=0.49\textwidth]{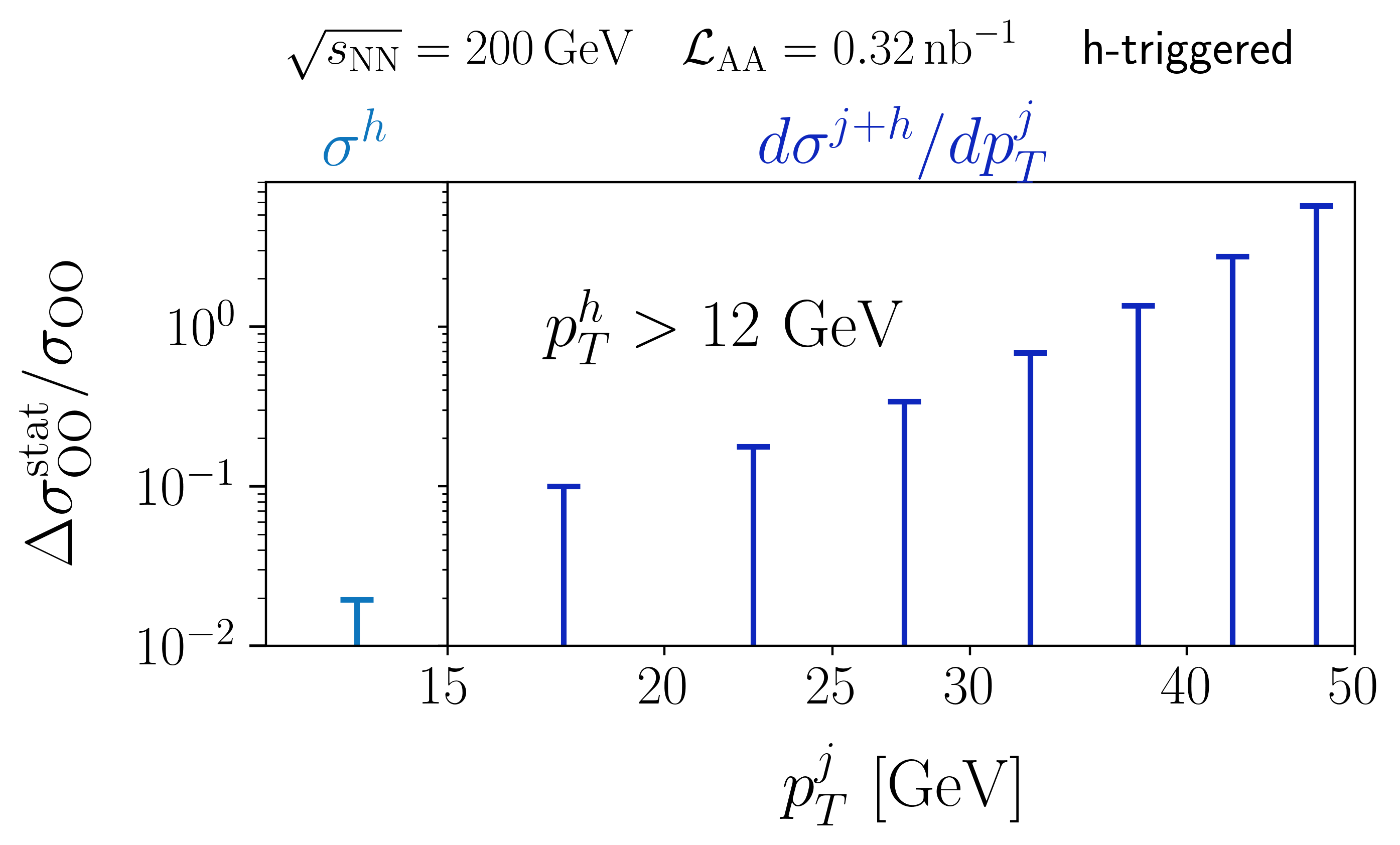}
    \caption{Projection of relative statistical uncertainties for trigger and coincidence cross sections at $\sqrt{s_\mathrm{NN}} = 200\,\mathrm{GeV}$ using cross sections obtained with EPPS21. \textit{Left}: Jet-triggered measurement with trigger momentum cutoff $p_{T,\mathrm{min}}^j=15\,\mathrm{GeV}$. Results are shown for trigger cross section $\sigma^j$ and away-side $\dd\sigma^{h+j}/\dd p_T$. \textit{Right}: Hadron-triggered measurement with $p_{T,\mathrm{min}}^h=12\,\mathrm{GeV}$. Results are shown for trigger cross section $\sigma^h$ and $\dd\sigma^{j+h}/\dd p_T^j$.} 
    \label{fig:statuncstar}
\end{figure}

Corresponding projections for the already collected 400 million minimum bias OO events~\cite{Liu:2022jtl} at RHIC are presented in \cref{fig:statuncstar}. Relative errors range from several percent to over 100\%, depending on the kinematics. This shows that the available statistics might not be sufficient for a precise measurement of these observables.

\bibliographystyle{JHEP}
\bibliography{master.bib}

\end{document}